\documentclass[aps,10pt,prd,notitlepage,showpacs,nofootinbib,superscriptaddress,compressed,floatfix]{revtex4-2}

\pdfoutput=1 % if your are submitting a pdflatex (i.e. if you have
\usepackage[utf8]{inputenc}
\usepackage[english]{babel}
\usepackage{amsmath}
\usepackage{graphicx}
\usepackage{dcolumn}
\usepackage{pbox}
\usepackage{amssymb}
\usepackage{pifont}
\usepackage{epsfig}
\usepackage[dvipsnames]{xcolor}
\usepackage{slashed}
\usepackage{amssymb}
\usepackage{mathrsfs}
\usepackage{color}
\usepackage[normalem]{ulem}
\usepackage{multirow}
\usepackage[font=small]{caption}
\usepackage[font=small]{subcaption}
\usepackage{url}
\usepackage{braket}
\usepackage{hyperref}
\usepackage{fontawesome5}
\usepackage{booktabs}
\graphicspath{{pic/}}
\definecolor{DarkBlue}{rgb}{0.7, 0.4, 1}
\definecolor{Blue}{rgb}{0, 0.8, 0}
\definecolor{MyLightBlue}{rgb}{0.5,0.7,1.9}
\definecolor{MyGreen}{rgb}{0.0,0.2, 0.0}
\definecolor{MyBrickRed}{rgb}{0, 0.5, 0.2}
\RequirePackage{hyperref}
\hypersetup{colorlinks, citecolor=Blue,linkcolor=DarkBlue, urlcolor=Green}

\newcommand{\bea}{\begin{eqnarray}}
\newcommand{\eea}{\end{eqnarray}}

\makeatletter

\makeatletter
\renewcommand\@makecaption[2]{%
  \par
  \vskip\abovecaptionskip
  \begingroup

   \small\rmfamily
    \begingroup
     \samepage
     \flushing
     \let\footnote\@footnotemark@gobble
     \@make@capt@title{#1}{#2}\par
    \endgroup
  \endgroup
  \vskip\belowcaptionskip
}
\makeatother

\DeclareUnicodeCharacter{2212}{-}
\newcommand {\black} {\color{black}}

\def\21{\mathrm{$SU(2)_L \otimes U(1)_Y$}}

\begin{document}
\title{Quantum spin correlations in \(Z^\prime\)-mediated \(t\bar t\) production at future lepton colliders}
\author{ShivaSankar K.A.}\email{a-shiva@particle.sci.hokudai.ac.jp}
\affiliation{Department of Physics, Hokkaido University, Sapporo 060-0810, Japan}
\author{Arindam Das}
\email{adas@particle.sci.hokudai.ac.jp}
\affiliation{Institute for the Advancement of Higher Education, Hokkaido University, Sapporo 060-0817, Japan}
\affiliation{Department of Physics, Hokkaido University, Sapporo 060-0810, Japan}
\author{Sanjoy Mandal}
\email{smandal@kias.re.kr}
\affiliation{Korea Institute for Advanced Study, Seoul 02455, Korea}
%\date{\today}
%%%%%%%%%%%%%%%%%%%%%%%%%%%%%%%%%%%%%%%%%%%%%%%%%%%%%%%%%%%%%%%%%%%%%%%%%%%%%%%%%%%
\begin{abstract}
We study quantum spin correlations in top-quark pair production at future lepton colliders in the presence of a neutral gauge boson from anomaly-free general $U(1)$ extensions of the Standard Model.  The process $\ell^+\ell^-\to t\bar t$, with $\ell=e,\mu$, is analyzed through the spin-density matrix including $\gamma$, $Z$ and $Z^\prime$ exchange and their interference.  We focus on quantum-information observables such as the sufficient entanglement marker $\mathcal{D}_{\min}$, concurrence, purity and the maximal Clauser-Horne-Shimony-Holt (CHSH) parameter, and compare their behavior with conventional rate information. Within the $U(1)_X$ framework, we consider several representative charge assignments to investigate how different chiral structures influence these observables, with particular emphasis on the $Z^\prime$ resonance region and polarized $e^-e^+$ collisions, where the two allowed initial-state helicity configurations can be selectively enhanced.  We show that electron-beam polarization provides a direct handle on the left- and right-handed lepton charges of various $U(1)_X$ scenarios. These results demonstrate that quantum spin observables provide information complementary to cross sections and angular distributions in searches for chiral neutral gauge interactions.
\end{abstract}
%%%%%%%%%%%%%%%%%%%%%%%%%%%%%%%%%%%%%%%%%%%%%%%%%%%%%%%%%%%%%%%%%%%%%%%%%%%%%%%%%%%%%
\maketitle

\section{Introduction}
%%%%%%%%%%%%%%%%%%%%%
Top-quark pair production is a clean laboratory for studying the quantum structure of high-energy scattering.  Since the top quark decays before hadronization, its spin information is preserved in its decay products, making the $t\bar t$ spin-density matrix experimentally accessible.  Measurements of top-quark spin correlations were first performed at the Tevatron by the CDF and D0 Collaborations \cite{CDF:2010yag,D0:2011rkb,D0:2015kta} and have subsequently been measured with remarkable precision at the LHC \cite{ATLAS:2012ao,ATLAS:2014aus,ATLAS:2014abv,ATLAS:2019zrq,ATLAS:2023fsd,CMS:2013roq,CMS:2015cal,CMS:2019nrx,CMS:2024zkc,CMS:2024pts}.  These measurements have motivated a broader program connecting collider observables with quantum-information concepts such as spin entanglement, Bell nonlocality, purity and other observables constructed from the spin-density matrix \cite{Bell:1964kc,Aspect:1981nv,Bloch:2012uep,Aguilar-Saavedra:2022uye,Afik:2020onf,Severi:2021cnj,Afik:2022kwm,Maltoni:2024tul,Maltoni:2024csn,Han:2023fci,Severi:2022qjy,Fabbrichesi:2021npl,Aoude:2022imd,Durupt:2025wuk,Jeans:2026eys,Xie:2026ixr,Barr:2022wyq,Fabbrichesi:2022ovb,Pei:2026pba,Goncalves:2026nnx,Barr:2021zcp,Fabbrichesi:2023cev,Altakach:2026fpl,Cheng:2024btk,Han:2024ugl,Guo:2026yhz,Arai:2026jtc,Ehataht:2023zzt,Han:2025ewp,Yang:2026uwu,Cheng:2024rxi,Zhang:2026nwm,Altakach:2022ywa,Afik:2025ejh,DelGratta:2025qyp,Liu:2026gxj,Lee:2026bvm}.

Future lepton colliders operating above the top-pair threshold, $\sqrt{s}>2m_t$, such as the ILC, CLIC, FCC-ee, CEPC and future muon colliders \cite{Linssen:2012hp,Behnke:2013xla,Accettura:2023ked,Aime:2022flm,InternationalMuonCollider:2024jyv,CEPCPhysicsStudyGroup:2022uwl,Ai:2025cpj,Blondel:2025kjz,Blondel:2019jmp}, provide a particularly favorable environment for this program.  In contrast to hadron colliders, the initial state is colorless and kinematically well constrained.  Proposed $e^-e^+$ machines cover benchmark energies from the top threshold and Higgs/top factories up to the multi-TeV regime, with CLIC motivating a representative $\sqrt{s}=3$ TeV polarized-beam study, while future muon colliders can extend the unpolarized reach to even higher energies.  The process $\ell^+\ell^-\to t\bar t$ therefore gives direct access to the helicity amplitudes that determine the final-state spin density matrix.  Conventional observables such as total rates, invariant-mass spectra, angular distributions and asymmetries remain essential, but they do not exhaust the available information.  Quantum observables probe how the scattering process populates and coherently mixes the $t\bar t$ spin states, and are therefore directly sensitive to the chiral structure of the underlying interaction.

Neutral gauge bosons provide a natural target for such spin-based probes.  A useful reference point is the gauged $B-L$ model \cite{Davidson:1978pm,Davidson:1979wr,Marshak:1979fm,Mohapatra:1980qe,Wetterich:1981bx,Masiero:1982fi,Mohapatra:1982xz,Buchmuller:1991ce}, which connects a new Abelian gauge symmetry with the origin of neutrino masses and flavor mixing \cite{ParticleDataGroup:2024cfk}, for instance through the type-I seesaw mechanism \cite{Minkowski:1977sc,Yanagida:1979as,Yanagida:1980xy,Gell-Mann:1979vob,Mohapatra:1979ia,Schechter:1980gr}.  More general anomaly-free $U(1)$ extensions retain the same minimal ingredients, including three right-handed neutrinos (RHNs), but allow the left- and right-handed SM fermions to carry different charges.  The corresponding $Z^\prime$ interactions are then chiral, with vector and axial-vector components that can modify both the rate and the spin structure of fermion-pair production \cite{Appelquist:2002mw,Coriano:2014mpa,Das:2017flq,Das:2017deo,Das:2021esm}.  These scenarios can be explored at the LHC, future hadron colliders \cite{FCC:2018bvk,FCC:2018byv,FCC:2018vvp,FCC:2025lpp,FCC:2025uan}, and future lepton colliders \cite{Benedikt:2025kwi,MuonCollider:2022ded,MuonCollider:2022nsa,MuonCollider:2022xlm}.

The key point for the present work is that a chiral $Z^\prime$ changes the helicity amplitudes for $\ell^+\ell^-\to t\bar t$.  Since the spin-density matrix is built from these amplitudes, the resulting quantum-information observables can distinguish effects that may look similar in unpolarized rates.  For $e^-e^+$ colliders, beam polarization sharpens this handle: for massless initial leptons only the opposite-helicity channels contribute, and polarized electron and positron beams can vary their relative weights.  We do not assume polarized muon beams, for which no comparable collider program exists.  The polarized part of our analysis should therefore be read as an electron-positron collider study of the left- and right-handed lepton charges of the $Z^\prime$.

In this work, we investigate three representative anomaly-free general $U(1)$ scenarios, namely the $U(1)_X$, $U(1)_{xq-\tau_R^3}$, and $U(1)_{q+\tilde{x}u}$ models \cite{Das:2017flq,Carena:2004xs,Hashimoto:2014ela}.  We study the processes $e^+e^- \rightarrow t\bar{t}$ and $\mu^+\mu^- \rightarrow t\bar{t}$ at future lepton colliders with center-of-mass energies ranging from the top-pair threshold to the multi-TeV regime.  The production proceeds through $s$-channel exchange of the photon, the SM $Z$ boson and the $Z^\prime$, including all interference effects.  We construct the spin-density matrix and evaluate the sufficient entanglement marker $\mathcal{D}_{\min}$, concurrence, purity and the maximal CHSH parameter.  We also study polarized $e^-e^+$ beams in detail, using the $U(1)_X$ model as the representative case and $\sqrt{s}=3$ TeV as the high-energy electron-collider benchmark, and compare the quantum-observable response with the corresponding differential rate information.

This paper is organized as follows. In Sec.~\ref{sec:model}, we introduce the general $U(1)$ scenarios considered in this work. In Sec.~\ref{sec:secIII}, we review the density-matrix formalism and define the quantum-information observables employed in our analysis. In Sec.~\ref{helicity}, we present the helicity amplitudes for top-quark pair production within both the SM and the general $U(1)$ frameworks. Our numerical results and phenomenological discussions are given in Sec.~\ref{RD}. Finally, our conclusions are presented in Sec.~\ref{sec:conc}.

%%%%%%%%%%%%%%%%%%%%%%%%%%%%%%%%%%%%%%%%%%
\section{General $U(1)$ extensions of the Standard Model}
\label{sec:model}
%%%%%%%%%%%%%%%%%%%%%%%%%
We consider anomaly-free general $U(1)$ extensions of the SM with three generations of SM-singlet RHNs and one SM-singlet scalar field.  The RHNs cancel gauge and mixed gauge-gravitational anomalies and can generate light neutrino masses through the type-I seesaw mechanism, while the singlet scalar breaks the extra gauge symmetry and gives mass to the neutral gauge boson $Z^\prime$.  For the present analysis, the relevant model information is the chiral charge assignment of the SM fermions and the resulting $Z^\prime$ couplings entering $\ell^+\ell^-\to t\bar t$.

\iffalse
The Yukawa interactions consistent with the $\text{SM} \otimes U(1)$ gauge symmetry are given by
\begin{align}
&{\cal L}^{\rm Yukawa} = - Y_u^{\alpha \beta} \overline{q_L^\alpha} H u_R^\beta
- Y_d^{\alpha \beta} \overline{q_L^\alpha} \tilde{H} d_R^\beta
- Y_e^{\alpha \beta} \overline{\ell_L^\alpha} \tilde{H} e_R^\beta \nonumber\\
&- Y_\nu^{\alpha \beta} \overline{\ell_L^\alpha} H N_R^\beta
-\frac{1}{2}Y_N^\alpha \Phi\,\overline{(N_R^\alpha)^c}N_R^\alpha
+{\rm h.c.},
\label{LYk}
\end{align}
where $\tilde{H}=i\tau^2H^*$, with $\tau^2$ denoting the second Pauli matrix. The spontaneous breaking of the general $U(1)$ symmetry via the VEV of the singlet scalar $\Phi$ generates Majorana masses for the RHNs and simultaneously gives mass to the neutral gauge boson, $Z^\prime$. Subsequently, electroweak symmetry breaking induces the Dirac masses of the neutrinos, and the interplay between the Dirac and Majorana mass terms gives rise to the observed tiny neutrino masses and flavor mixing through the seesaw mechanism.
\fi

The three representative classes used in this work are summarized in Tab.~\ref{tab1}.  The parameters $x_H,x_\Phi,x$ and $\tilde{x}$ label different anomaly-free charge assignments.  Different points in this parameter space may lead to identical SM-fermion charges and therefore identical predictions for the $t\bar t$ observables considered below.
%%%%%%%%%%%%%%%%%%%%%%%%%%%%%%%%%%%%%%%%%%%%%%%%%%%%%%%%%%%%
\begin{table}[!htbp]
	\begin{center}
	\resizebox{\textwidth}{!}{%
		\begin{tabular}{| c| c || c | c |c||c|}
			\hline
			\hspace{0.5cm}Fields \hspace{0.5cm}   & \hspace{0.5cm} $SU(3)_c\otimes SU(2)_L\otimes U(1)_Y$ \hspace{0.5cm} & \hspace{0.5cm} $U(1)_X$ \hspace{0.5cm} &\hspace{0.5cm} $U(1)_{xq-\tau_R^3}$ \hspace{0.5cm}&\hspace{0.5cm} $U(1)_{q+ \tilde{x} u}$ \hspace{0.5cm}\\
			\hline \hline
			$q_L^i$ \ \             & $(3, 2, \frac{1}{6})$\ \      & $x_q= \frac{1}{6}x_H + \frac{1}{3} x_{\Phi}$ & $x$ & $\frac{1}{3}$ \\[0.1cm]
            $u_R^i$ \ \             & $(3, 1,  \frac{2}{3})$\ \      & $x_u=\frac{2}{3}x_H + \frac{1}{3}x_{\Phi} $& $-1+4 x$ & $\frac{\tilde{x}}{3}$ \\[0.1cm]
            $d_R^i$ \ \             & $(3, 1, -\frac{1}{3})$\ \      & $x_d=-\frac{1}{3} x_H + \frac{1}{3} x_{\Phi} $ & $1-2x $ & $\frac{2-\tilde{x}}{3} $ \\[0.1cm]
            \hline \hline
			$\ell_L^i$ \ \             & $(1, 2, -\frac{1}{2})$\ \         &$x_\ell=-\frac{1}{2} x_H - x_{\Phi}$ &$-3x$ &$-1$ \\[0.1cm]
			$ e_R^i$ \ \        & $(1, 1, -1)$\ \                  & $x_e=-x_H - x_{\Phi}$ &$1-6x$ &$-(\frac{2+\tilde{x}}{3})$ \\[0.1cm]
            \hline \hline
			$H$       \ \  & $(1, 2, -\frac{1}{2})$\ \               & $-\frac{1}{2} x_H$ & $1-3x$ & $\frac{1-\tilde{x}}{3}$ \\[0.1cm]
			\hline \hline
			$ N^j$ \ \           & $(1,1,0)$ \ \ 					    &$x_N=-x_{\Phi}$ &$-1$ &$\frac{-4+\tilde{x}}{3}$ \\[0.1cm]
			$\Phi$			\ \ &$(1,1,0)$ 				  \ \	    &$2\;x_{\Phi}$ &$2$ &$-2 (\frac{-4+\tilde{x}}{3})$ \\
			\hline \hline
		\end{tabular}%
	}
	\end{center}
	\caption{Fields with the charge assignments of minimal $U(1)$ extensions of the SM where \(i\) and \(j\) are generation indices for the SM fermions and RHNs, respectively. Here general $U(1)$ charges involving $x_H$, $x_{\Phi}$, $x$ and $\tilde{x}$ are free real parameters.}
	\label{tab1}
\end{table}
%%%%%%%%%%%%%%%%%%%%%%%%%%%%%%%%%%%%%%%%

For the phenomenological analysis, we fix $x_\Phi=1$ in the $U(1)_X$ model and use representative charge assignments that either reproduce familiar limits or emphasize chiral effects.  The choice $x_H=-2$ gives the $U(1)_R$ scenario, in which the SM quark and lepton doublets are neutral under $U(1)_X$ while the corresponding right-handed fields remain charged; the same SM-fermion charge assignment is obtained in the $U(1)_{xq-\tau_R^3}$ model for $x=0$, but it has no equivalent realization in $U(1)_{q+\tilde{x}u}$.  The benchmark $x_H=-0.5$ corresponds to $x=1/4$ and $\tilde{x}=0$ in the other two models and gives a complementary chiral pattern in which the right-handed up-type quarks are neutral.  The vector-like B$-$L limit is recovered for $x_H=0$ in $U(1)_X$, $x=1/3$ in $U(1)_{xq-\tau_R^3}$ and $\tilde{x}=1$ in $U(1)_{q+\tilde{x}u}$; in this case left- and right-handed fermions of the same species carry identical charges.  We also consider $x_H=2$, equivalent to $x=2/3$ in $U(1)_{xq-\tau_R^3}$, which has no realization in $U(1)_{q+\tilde{x}u}$.  These benchmarks span vector-like and chiral $Z^\prime$ interactions and are therefore well suited for testing how spin-density-matrix observables respond to the underlying gauge structure.

The interaction of the $Z^\prime$ with SM fermions is given by
\bea
\mathcal{L} = -g_X (\overline{f_L}\gamma^\mu q_{f_{L}^{}}^{}  f_L+ \overline{f_R}\gamma^\mu q_{f_{R}^{}}^{}  f_R) Z_\mu^\prime.
\label{Lag1}
\eea
where $q_{f_L}$ and $q_{f_R}$ are the corresponding general $U(1)$ charges of the left- and right-handed fermions listed in Tab.~\ref{tab1}.  This chiral structure enters directly into the helicity amplitudes for $\ell^+\ell^-\to t\bar t$.  The same charges determine the $Z^\prime$ total width used in the propagator.  For a charged-lepton or quark final state, the partial width is
\begin{align}
\label{eq:width-ll}
\Gamma(Z^\prime \to \bar{f} f)
= \;N_C^{} \frac{M_{Z^\prime}^{} g_{X}^2}{24 \pi} \left[ \left( q_{f_L^{}}^2 + q_{f_R^{}}^2 \right) \left( 1 - \frac{m_f^2}{M_{Z^\prime}^2} \right) \right. \left. + \,6 q_{f_L^{}}^{} q_{f_R^{}}^{} \frac{m_f^2}{M_{Z^\prime}^2} \right]
\sqrt{1-\frac{4 m_f^2}{M_{Z^\prime}^2}}~,
\end{align}
where $m_f$ is the fermion mass and $N_C=1~(3)$ for SM leptons (quarks).  Neglecting light-neutrino masses, the partial width into one generation of light neutrinos is
\begin{align}
\label{eq:width-nunu}
    \Gamma(Z^\prime \to \nu \nu)
    =  \frac{M_{Z^\prime}^{} g_{X}^2}{24 \pi} q_{\ell_L}^2~,
\end{align}
where $q_{\ell_L}$ is the charge of the SM lepton doublet.  This contribution is multiplied by three when all light-neutrino generations are included.  We take the heavy neutrinos to satisfy $M_N>M_{Z^\prime}/2$, so the $Z^\prime\to NN$ channel is closed and is not included in the widths used for the $t\bar t$ analysis.
\iffalse
In general $U(1)$ extended SM scenarios, $Z^\prime$ interacts with heavy neutrinos directly due to the nonzero general $U(1)$ charges of the RHNs. The interaction Lagrangian can be written in the following way
\bea
\mathcal{L}_N= -\frac{1}{2}g_X q_{N_R} \overline{N} \gamma_\mu \gamma_5 N Z_{\mu}^\prime,
\label{neut}
\eea
where $q_{N_R}$ is the  general $U(1)$ charge of heavy neutrinos. The partial decay width of $Z^\prime$ into a pair of single generation heavy neutrino can be written as
\begin{align}
\label{eq:width-NN}
    \Gamma(Z^\prime \to N^\alpha N^\alpha)
    = \frac{ g_{X}^2 M_{Z^\prime}^{}}  {24 \pi} q_{N_R^{}}^2 \left( 1 - \frac{4 M_{N_\alpha}^2}{M_{Z^\prime}^2} \right)^{\frac{3}{2}}~,
\end{align}
where $M_{N_\alpha}$ being the heavy neutrino mass and Eq.~(\ref{eq:width-NN}) will be multiplied by $3$ for when $\Gamma(Z^\prime \to N^\alpha N^\alpha)$ will be calcullated for $3$ generations of the heavy neutrinos. As we consider $M_N > M_{Z^\prime}/2$, this mode will be forbidden in our work.
\fi
%%%%%%%%%%%%%%%%%%%%%%%%%%%%%%%%%%%%%%%%%%%%%%%%%%%%%%%%%%%%%%%%%%%%%%%%%%%%%%%%%%%%%%%%%%%
\section{Quantum observables at high-energy lepton colliders}
\label{sec:secIII}
%%%%%%%%%%%%%%%%%%%%%%%%%%%%%%%%%%%%%%%%%%%%%%%%%%%%%%%%%%%%%%%%%%%%%%%%%%%
Our calculation begins with helicity amplitudes and ends with scalar diagnostics of the produced spin state.  We describe that chain in this order so that the normalization of each quantity remains explicit.  For
$\ell^-(\lambda_{\ell^-})\ell^+(\lambda_{\ell^+})\to
t(\lambda_t)\bar t(\lambda_{\bar t})$, with $\ell=e$ or $\mu$, we abbreviate the amplitude as
\begin{equation}
\mathcal{M}_{\lambda_-\lambda_+}^{\lambda_t\lambda_{\bar t}}
\equiv
\mathcal{M}(\lambda_{\ell^-},\lambda_{\ell^+};
\lambda_t,\lambda_{\bar t}) .
\end{equation}
Here and below the helicity symbols $\pm$ stand for $\pm1/2$.  Before normalizing the final state, we retain its event-weight information in the production matrix.  For unpolarized beams it is
\begin{align}
\mathcal{R}_{\lambda_t\lambda_{\bar t},
\lambda'_t\lambda'_{\bar t}}
&=
\frac{1}{4}
\sum_{\lambda_-,\lambda_+}
\mathcal{M}_{\lambda_-\lambda_+}^{\lambda_t\lambda_{\bar t}}
\left(
\mathcal{M}_{\lambda_-\lambda_+}^{\lambda'_t\lambda'_{\bar t}}
\right)^\ast ,
\label{spinmat}
\end{align}
where the two final-state helicity pairs specify the row and column of $\mathcal R$.  Removing its overall normalization gives
\begin{equation}
\rho_{\lambda_t\lambda_{\bar t},\lambda'_t\lambda'_{\bar t}}
=\frac{\mathcal R_{\lambda_t\lambda_{\bar t},\lambda'_t\lambda'_{\bar t}}}
{\operatorname{Tr}\mathcal R},
\label{spin-d1}
\end{equation}
which has unit trace and contains only information about the spin composition of the selected events.

Polarized beams change the statistical weight assigned to each incoming electron and positron helicity.  We therefore replace the factor $1/4$ in Eq.~(\ref{spinmat}) by
\begin{align}
\mathcal{R}_{\lambda_t\lambda_{\bar t},
\lambda'_t\lambda'_{\bar t}}(P_{e^-},P_{e^+})
=
\sum_{\lambda_-,\lambda_+}
w_{e^-}(\lambda_-;P_{e^-})
w_{e^+}(\lambda_+;P_{e^+})
%\nonumber\\
%\hspace{0.4cm}
\times \mathcal{M}_{\lambda_-\lambda_+}^{\lambda_t\lambda_{\bar t}}
\left(
\mathcal{M}_{\lambda_-\lambda_+}^{\lambda'_t\lambda'_{\bar t}}
\right)^\ast ,
\label{spinmat-pol}
\end{align}
with
\begin{equation}
w_{e^\mp}(\lambda;P_{e^\mp})=\frac{1+\lambda P_{e^\mp}}{2}, \,\,\,\,\, \lambda=\pm 1.
\label{polweight}
\end{equation}
The sign label $\lambda=\pm1$ in Eq.~(\ref{polweight}) is twice the physical helicity, whereas $P_{e^\mp}\in[-1,1]$ is the polarization degree of the corresponding beam.  Thus $P=+1$ selects a fully right-handed beam, $P=-1$ selects a fully left-handed beam, and $P=0$ gives an equal mixture.  We apply this weighting only to the $e^-e^+$ analysis; all muon-collider results presented here are unpolarized.

Neglecting the electron mass leaves only the $e^-_R e^+_L$ and $e^-_L e^+_R$ initial states for the neutral vector currents used here. It is useful to separate the relative mixture of the surviving channels from their common normalization.  Defining
\begin{align}
W_{RL}
&=
w_{e^-}(+;P_{e^-})
w_{e^+}(-;P_{e^+}), \\
W_{LR}
&=
w_{e^-}(-;P_{e^-})
w_{e^+}(+;P_{e^+}), \\
\sin^2\Phi
&\equiv
\frac{W_{RL}}{W_{RL}+W_{LR}}
%\nonumber\\
=
\frac{(1+P_{e^-})(1-P_{e^+})}
{2(1-P_{e^-}P_{e^+})}.
\label{sinphi}
\end{align}

The endpoints $\sin^2\Phi=1$ and $0$ select the $RL$ and $LR$ channels, respectively. This parametrization follows the polarized-beam construction of Ref.~\cite{Altakach:2026fpl}. For fixed $\Phi$, the common factor $W_{RL}+W_{LR}=(1-P_{e^-}P_{e^+})/2$ cancels when $\mathcal R$ is normalized to obtain $\rho$. Consequently, $\sin^2\Phi$ uniquely specifies the normalized spin observables, but it does not by itself specify the polarized cross section. We therefore use the $\sin^2\Phi$ parametrization alone only for normalized-observable maps. The distinction between $\mathcal R$ and $\rho$ is also the distinction between yield and state composition. Indeed, summing the diagonal entries of $\mathcal R$ gives
\begin{align}
{\rm Tr}\,\mathcal{R}(P_{e^-},P_{e^+})
&=
\sum_{\lambda_t,\lambda_{\bar t}}
\mathcal{R}_{\lambda_t\lambda_{\bar t},
\lambda_t\lambda_{\bar t}}(P_{e^-},P_{e^+})
\nonumber\\
&=
\sum_{\lambda_-,\lambda_+}
w_{e^-}(\lambda_-;P_{e^-})
w_{e^+}(\lambda_+;P_{e^+})
%\nonumber\\
%&\hspace{0.35cm}
\times \sum_{\lambda_t,\lambda_{\bar t}}
\left|
\mathcal{M}_{\lambda_-\lambda_+}^{\lambda_t\lambda_{\bar t}}
\right|^2 .
\label{trace-msq}
\end{align}
This trace is the polarization-weighted squared amplitude summed over final spins. The differential rate is therefore proportional to it,
\begin{equation}
\frac{d\sigma(P_{e^-},P_{e^+})}{d\Omega}
\propto
{\rm Tr}\,\mathcal{R}(P_{e^-},P_{e^+}) .
\label{dsig-trR}
\end{equation}
For an explicitly specified polarization pair, the polarized-to-unpolarized differential rate ratio can consequently be evaluated directly from traces,
\begin{align}
\mathcal{R}_{\sigma}(P_{e^-},P_{e^+};\sqrt{s},\theta)
&\equiv
\frac{d\sigma(P_{e^-},P_{e^+})/d\Omega}
{d\sigma(0,0)/d\Omega}
%\nonumber\\
=
\frac{{\rm Tr}\,\mathcal{R}(P_{e^-},P_{e^+})}
{{\rm Tr}\,\mathcal{R}(0,0)}.
\label{rate-ratio}
\end{align}
The common proportionality factor cancels between numerator and denominator. For comparisons at a common $(\sqrt{s},\theta,\sin^2\Phi)$ point, we use the double ratio
\bea
\Delta \mathcal{O}\equiv \mathcal{O}_{U(1)_X}-\mathcal{O}_{\rm SM},
\qquad
\Delta_{\rm rate}\equiv
\log_{10}\left[
\frac{\mathcal{R}_{\sigma}^{U(1)_X}}
{\mathcal{R}_{\sigma}^{\rm SM}}
\right],
\label{delta-defs}
\eea
where $\mathcal O$ is any normalized spin observable. The common factor $W_{RL}+W_{LR}$ cancels in the ratio $\mathcal R_{\sigma}^{U(1)_X}/\mathcal R_{\sigma}^{\rm SM}$, so $\Delta_{\rm rate}$ is fixed by $\sin^2\Phi$ even though either polarized cross section separately is not. The logarithm treats enhancement and suppression symmetrically on the color scale.

We next translate $\rho$ into quantities with a direct spin interpretation. A general $4\times4$ Hermitian matrix has 16 real parameters, and unit normalization removes one; hence any normalized two-qubit state has 15 independent real parameters. We organize these parameters as two three-component polarization vectors,
\begin{equation}
\mathcal B_1=\{\mathcal B_{1i}\},\qquad
\mathcal B_2=\{\mathcal B_{2i}\},
\label{cor2}
\end{equation}
and a $3\times3$ correlation matrix
\begin{equation}
\mathcal C=\{C_{ij}\},\qquad i,j=1,2,3.
\label{cor1}
\end{equation}
Their Pauli-basis expansion is
\begin{align}
\rho=\frac14\bigg[&\mathbf1\otimes\mathbf1
+\sum_i\mathcal B_{1i}\sigma_i\otimes\mathbf1
+\sum_j\mathcal B_{2j}\mathbf1\otimes\sigma_j
+\sum_{i,j}C_{ij}\sigma_i\otimes\sigma_j\bigg].
\label{densitym}
\end{align}
Orthogonality of the Pauli matrices then gives the coefficients without any model-dependent assumptions:
\bea
\mathcal{B}_{1i}&=& {\rm Tr}\!\left[\rho\, (\sigma_i \otimes \mathbf{1})\right],  \\
\mathcal{B}_{2j}&=& {\rm Tr}\!\left[\rho\, (\mathbf{1} \otimes \sigma_j)\right], \\
C_{ij}&=& {\rm Tr}\!\left[\rho\, (\sigma_i \otimes \sigma_j)\right].
\label{coeffs}
\eea

\begin{figure}[!htbp]
\centering
\includegraphics[width=0.58\textwidth]{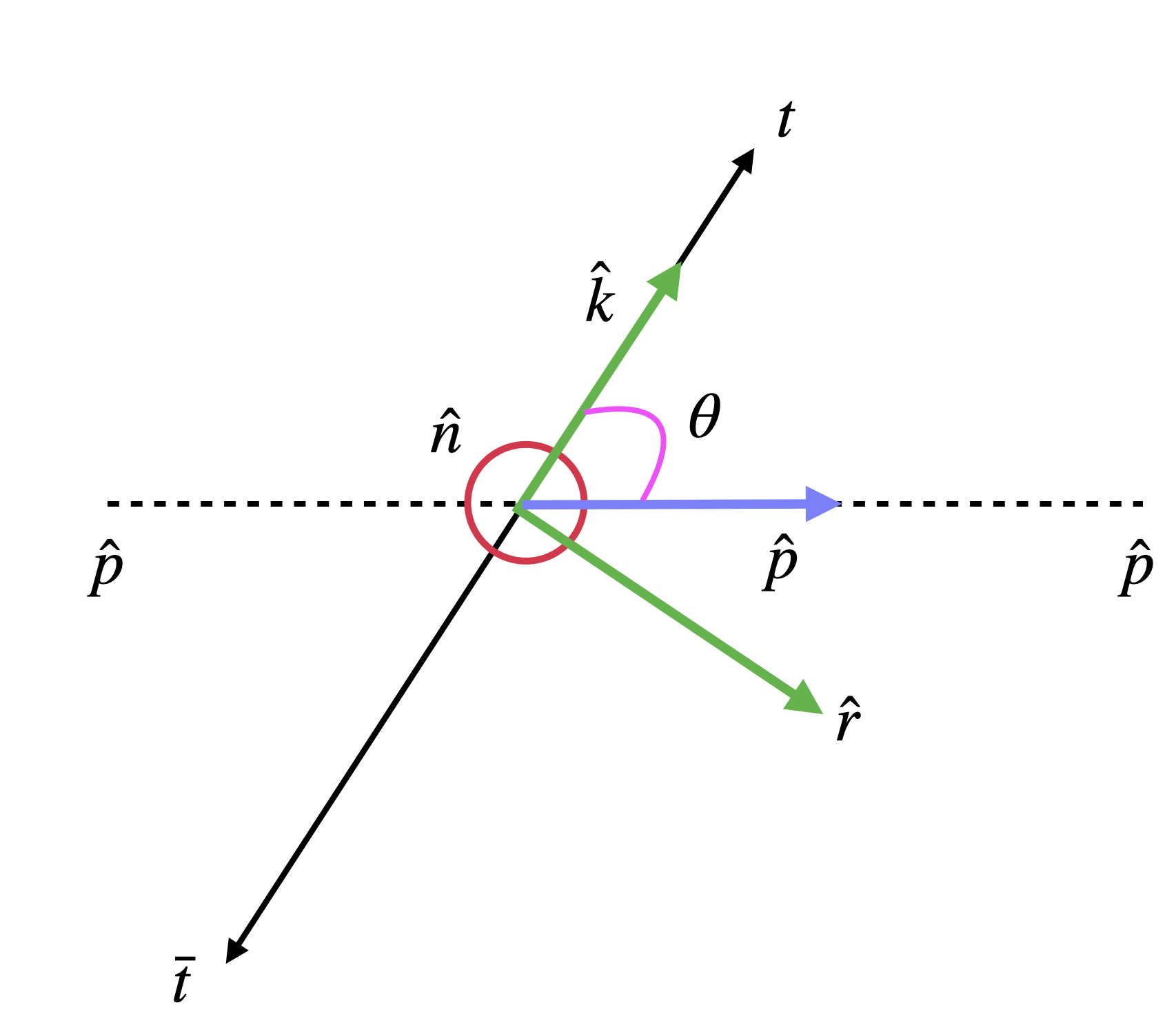}
\caption{Spin basis used for the $t\bar t$ density matrix in the pair rest frame.  The beam and top directions define the scattering plane, while $\hat n$ is normal to it.}
\label{vec}
\end{figure}

The indices in Eqs.~(\ref{cor1})--(\ref{coeffs}) refer to the event-by-event triad shown in Fig.~\ref{vec}.  In the $t\bar t$ rest frame, let $\hat p$ point along the incoming negatively charged lepton and $\hat k$ along the top momentum.  We define the orthonormal basis with
\begin{equation}
\hat r=\frac{\hat p-\hat k\cos\theta}{\sin\theta},
\qquad
\hat n=\frac{\hat p\times\hat k}{\sin\theta},
\qquad
\cos\theta=\hat p\cdot\hat k.
\end{equation}
For the explicit matrix representation, we quantize both spins along $\hat k$ and order the states as
\bea
|1\rangle = |+,-\rangle,\,\,
|2\rangle = |+,+\rangle,\,\,
|3\rangle = |-,-\rangle,\,\,
|4\rangle = |-,+\rangle,
\eea
where the entries label top and antitop helicity.  Because the antitop moves opposite to $\hat k$, its helicity sign is opposite to its spin projection on the common $\hat k$ axis.

We use four complementary functions of $\rho$.  The simplest is the purity,
\begin{equation}
\Gamma={\rm Tr}(\rho^2),
\label{purity}
\end{equation}
which ranges from $1/4$ for the maximally mixed two-qubit state to unity for a pure state.  In the polarized analysis it indicates whether changing the initial helicity mixture merely changes the yield or also changes how coherently the final spin states are populated.

Following Refs.~\cite{Bernreuther:2004jv,Bernreuther:2010ny,Bernreuther:2013aga,Bernreuther:2015yna,Maltoni:2024tul}, we construct four entanglement markers from the three diagonal spin correlations:
\bea
\mathcal{D}^{(1)}&=& \frac{1}{3} \big(C_{kk}+C_{rr}+C_{nn}\big), \\
\mathcal{D}^{(k)}&=& \frac{1}{3} \big(C_{kk}-C_{rr}-C_{nn}\big), \\
\mathcal{D}^{(r)}&=& \frac{1}{3} \big(-C_{kk}+C_{rr}-C_{nn}\big),\\
\mathcal{D}^{(n)}&=& \frac{1}{3} \big(-C_{kk}-C_{rr}+C_{nn}\big),
\label{Dvar}
\eea
and retain their minimum $\mathcal D_{\min}=\min_a\mathcal D^{(a)}$.
A convenient sufficient criterion for entanglement is \cite{Peres:1996dw,Horodecki:1996nc}
\begin{equation}
\mathcal D_{\min}< -\frac13,
\label{suff2}
\end{equation}
Failure to cross this boundary is inconclusive because off-diagonal correlations and one-particle polarizations do not enter this criterion. This experimentally robust construction has played a central role in top-entanglement measurements at the LHC \cite{ATLAS:2023fsd,CMS:2024pts}. For an exact two-qubit entanglement measure we also calculate concurrence. Introduce the spin-flipped density matrix \cite{Hill:1997pfa,Wootters:1997id}
\begin{equation}
\widetilde\rho=(\sigma_2\otimes\sigma_2)\rho^*(\sigma_2\otimes\sigma_2),
\label{den-2}
\end{equation}
and the positive matrix
\begin{equation}
\mathcal R_\rho=\sqrt{\sqrt\rho\,\widetilde\rho\,\sqrt\rho}.
\label{den-3}
\end{equation}
If its eigenvalues are ordered as $\lambda_1\geq\lambda_2\geq\lambda_3\geq\lambda_4\geq0$, then
\begin{equation}
C(\rho)=\max\{0,\lambda_1-\lambda_2-\lambda_3-\lambda_4\}.
\end{equation}
Thus $C=0$ characterizes separable two-qubit states and $C=1$ a maximally entangled state. The associated entanglement of formation is \cite{Wootters:1997id}
\begin{equation}
E(\rho)=h\!\left(\frac{1+\sqrt{1-C(\rho)^2}}{2}\right), \text{ with } h(x)=-x\log_2x-(1-x)\log_2(1-x).
\label{conc}
\end{equation}
When top and antitop polarizations vanish ($\mathcal B_1=\mathcal B_2=0$) and $\mathcal C$ is diagonal, $\mathcal D_{\min}$ and concurrence are related by
\begin{equation}
C=\frac12\max\{0,-1-3\mathcal D_{\min}\}.
\label{conc2}
\end{equation}
Outside that restricted class of states, concurrence is evaluated from the full density matrix rather than inferred from $\mathcal D_{\min}$. Finally, we test whether the correlations are incompatible with a local hidden-variable description.  For spin measurements along unit vectors $\hat a$ and $\hat b$, define
\begin{equation}
E(\hat a,\hat b)
=\operatorname{Tr}\!\left[\rho(\hat a\cdot\boldsymbol\sigma)
\otimes(\hat b\cdot\boldsymbol\sigma)\right]
=\hat a^{T}\mathcal C\hat b.
\end{equation}
The CHSH combination built from two settings on each side obeys
\begin{equation}
|E(\hat a_1,\hat b_1)-E(\hat a_1,\hat b_2)
+E(\hat a_2,\hat b_1)+E(\hat a_2,\hat b_2)|\leq2
\label{hvar}
\end{equation}
in any local hidden-variable theory \cite{Bell:1964kc,Clauser:1969ny,Fine:1982lwu,Bell:1964fg}.  Optimization over the four directions can be done algebraically.  If $e_1\geq e_2\geq e_3$ are the eigenvalues of $\mathcal C^T\mathcal C$, the largest attainable value is \cite{Horodecki:1995nsk,James:2001klt,Chen:2013epa}
\begin{equation}
\mathcal B_{\rm CHSH}\equiv B_{\max}=2\sqrt{e_1+e_2}.
\end{equation}
We denote this Bell parameter by $\mathcal B_{\rm CHSH}$.  Bell-inequality violation occurs for $\mathcal B_{\rm CHSH}>2$, while quantum mechanics limits it to $2\sqrt2$ \cite{Cirelson:1980ry}.  Unlike concurrence, this quantity diagnoses Bell nonlocality rather than entanglement alone.  Both can be obtained after reconstructing the spin density matrix from top-decay angular distributions \cite{Aguilar-Saavedra:2022uye,Afik:2022kwm,Severi:2021cnj,Fabbrichesi:2021npl}.

Numerically, we construct $\mathcal R$ from the SM and $U(1)$ helicity amplitudes, normalize it according to Eq.~(\ref{spin-d1}), and apply Eqs.~(\ref{coeffs}) and (\ref{Dvar}) directly.  Concurrence is always computed from the full $4\times4$ matrix, while the maximal CHSH parameter is obtained from the two largest eigenvalues of $\mathcal C^T\mathcal C$. As an independent validation, we compared our analytically constructed \(t\bar t\) spin-density matrices and the resulting \(\mathcal D_{\min}\), concurrence, and \(\mathcal B_{\rm CHSH}\) values with event-by-event results from the automated MadGraph density-matrix implementation of Ref.~\cite{Durupt:2025wuk}. After accounting for the helicity-state phase convention, we find numerical agreement for both the density-matrix elements and the derived quantum observables.\footnote{To perform the validation we used the numerical code of the \href{https://cp3.irmp.ucl.ac.be/projects/feynrules/wiki/GeneralU1}{general U(1) model} \cite{Das:2021esm} from FeynRules.}.

%%%%%%%%%%%%%%%%%%%%%%%%%%%%%%%%%%%%%%%%%%%%%%%%%%%%%%%%%%%%%%%%
\section{Production of $t\bar{t}$ in lepton colliders}
\label{helicity}
%%%%%%%%%%%%%%%%%%%%%%%%%%%%%%%%%%%%%%%%%%%%%%%
\begin{figure}[!htbp]
\centering
\includegraphics[width=0.72\textwidth]{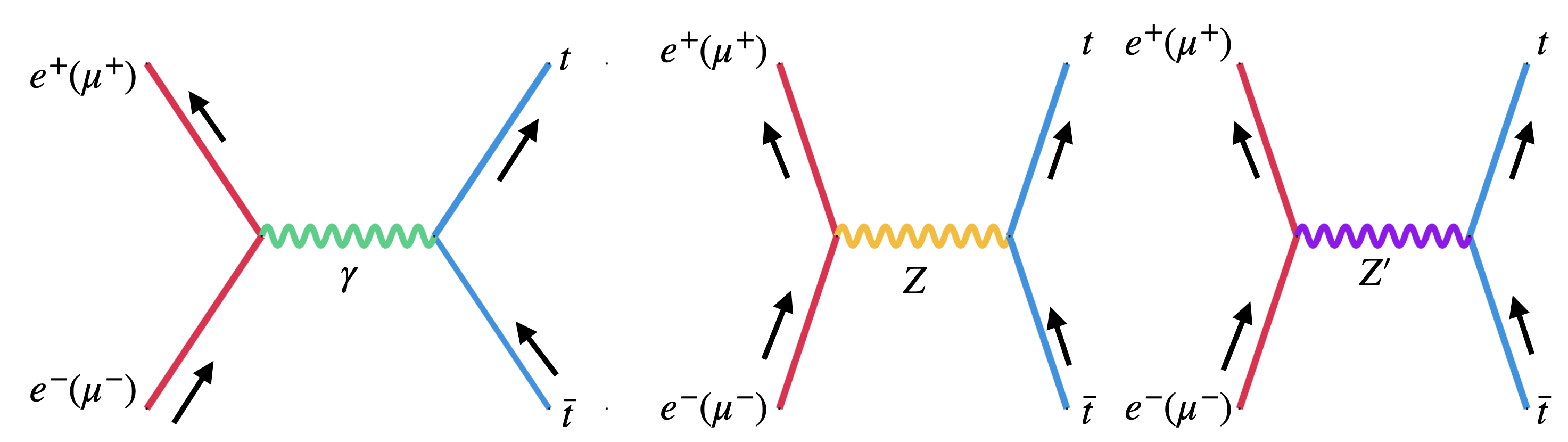}
\caption{Feynman diagrams for $t\bar t$ production at $e^-e^+$ and $\mu^-\mu^+$ colliders through $s$-channel exchange.  The first two diagrams show the SM photon and $Z$ contributions, while a general $U(1)$ extension adds $Z^\prime$ exchange.}
\label{FD1}
\end{figure}
%%%%%%%%%%%%%%%%%%%%%%%%%%%%%%%%%%%%%%%%%%%%%%%%%%%%%%%%%
We work in the center-of-mass (CM) frame with massless incoming leptons and an outgoing $t\bar t$ pair whose constituents have mass $m_t$. The four-momenta are
\bea
p_1^\mu=\frac{\sqrt{s}}{2}(1,0,0,1),\qquad
p_2^\mu=\frac{\sqrt{s}}{2}(1,0,0,-1),
\label{in-mom}
\eea
with the lepton (antilepton) along $+z$ ($-z$). Choosing the $xz$-plane as the scattering plane, the final-state momenta are
\begin{align}
k_1^\mu &=
\frac{\sqrt{s}}{2}
\left(1,\beta \sin\theta,0,\beta \cos\theta\right),\qquad
k_2^\mu =
\frac{\sqrt{s}}{2}
\left(1,-\beta \sin\theta,0,-\beta \cos\theta\right),
\label{out-mom}
\end{align}
with $\beta =\sqrt{1-4 m_t^2/s}$ the top velocity and $\theta$ the scattering angle of $t$ relative to the beam axis. The orthonormal spin-correlation basis is
\begin{align}
\hat k=\frac{\mathbf{k}_1}{|\mathbf{k}_1|},\, \, \, \,\,%&
\hat p=\frac{\mathbf p_1}{|\mathbf p_1|}, \, \, \, \, \, %\nonumber\\
\hat r=\frac{\hat p-\hat k\cos\theta}{\sin\theta},\,\,\, \, \, %&
\hat n=\frac{\hat p\times\hat k}{\sin\theta}.
\end{align}
This basis is used for the spin-correlation matrix $C_{ij}$ with
$i,j\in\{r,n,k\}$. We use the chiral representation of the gamma matrices,
\begin{align}
\gamma^0=
\begin{pmatrix}
0&\mathbf{I}_{2 \times 2}\\
\mathbf{I}_{2\times 2}&0
\end{pmatrix},
\;
\gamma^i=
\begin{pmatrix}
0&-\sigma^i\\
\sigma^i&0
\end{pmatrix},
%\nonumber
%\\
\gamma^5
=
\begin{pmatrix}
-\mathbf{I}_{2\times 2}&0\\
0&\mathbf{I}_{2\times 2}
\end{pmatrix},
\;
P_{L,R}=\frac{1\mp\gamma^5}{2},
\end{align}
and $\bar\psi=\psi^\dagger\gamma^0$.  For a unit three-vector $\hat p$ with
polar and azimuthal angles $(\theta,\phi)$, the two-component helicity spinors
are
\begin{equation}
\chi(\hat p)=
\begin{pmatrix}
\cos\frac{\theta}{2}\\[2pt]
e^{i\phi}\sin\frac{\theta}{2}
\end{pmatrix},
\qquad
\xi(\hat p)=
\begin{pmatrix}
-e^{-i\phi}\sin\frac{\theta}{2}\\[2pt]
\cos\frac{\theta}{2}
\end{pmatrix}
\end{equation}
obeying $(\boldsymbol{\sigma}\cdot\hat p)\chi(\hat p)=+\chi(\hat p)$ and
$(\boldsymbol{\sigma}\cdot\hat p)\xi(\hat p)=-\xi(\hat p)$, respectively.
For a fermion with $p^\mu=(E,\mathbf p)$ and $p=|\mathbf p|$, we use the
HELAS spinor convention \cite{Murayama:1992gi}
\small
\begin{align}
u(p,+) =
\begin{pmatrix}
\sqrt{E-p}\,\chi(\hat p)\\
\sqrt{E+p}\,\chi(\hat p)
\end{pmatrix},
%\nonumber\\
u(p,-)=
\begin{pmatrix}
\sqrt{E+p}\,\xi(\hat p)\\
\sqrt{E-p}\,\xi(\hat p)
\end{pmatrix},
%\nonumber\\
v(p,+) &=
\begin{pmatrix}
-\sqrt{E+p}\,\xi(\hat p)\\
\sqrt{E-p}\,\xi(\hat p)
\end{pmatrix},
%\nonumber\\
v(p,-) =
\begin{pmatrix}
\sqrt{E-p}\,\chi(\hat p)\\
-\sqrt{E+p}\,\chi(\hat p)
\end{pmatrix}.
\end{align}
\normalsize
The azimuthal dependence is absorbed into the orthonormal basis $\{\hat r,\hat n,\hat k\}$, so the amplitudes depend only on the polar angle $\theta$.  The full $t\bar t$ scattering amplitude is
\bea
\mathcal{M}(\lambda_{\ell^-}, \lambda_{\ell^+}; \lambda_{t}, \lambda_{\bar{t}}) = \mathcal{M}_{\rm SM}(\lambda_{\ell^-}, \lambda_{\ell^+}; \lambda_{t}, \lambda_{\bar{t}}) %\nonumber\\
+\mathcal{M}_{Z^\prime}(\lambda_{\ell^-}, \lambda_{\ell^+}; \lambda_{t}, \lambda_{\bar{t}})
\label{scat-1}
\eea
where $\mathcal M_{\rm SM}$ contains photon and $Z$ exchange, and $\mathcal M_{Z^\prime}$ is the contribution from the additional $U(1)$ gauge boson.  We denote helicities $-1/2$ and $+1/2$ by $-$ and $+$, respectively.  For a gauge boson $V\in\{Z,Z^\prime\}$ with mass $M_V$ and total width $\Gamma_V$, we define the propagator denominator $\Delta_V=s-M_V^2+iM_V\Gamma_V$.

%%%%%%%%%%%%%%%%%%%%%%%%%%%%%%%%%%%%%%%%%%%%%%%
%\subsection{Standard Model processes}
%%%%%%%%%%%%%%%%%%%%%%%%%%%%%%%%%%%%%%%%%%%%%%%%%%%%%%%%%%%%%%%%%%%%%%%%%%%%%%
The SM helicity amplitudes from $s$-channel photon and $Z$ exchange are
\begin{align}
\mathcal{M}_{\rm SM }(+,-;\pm,\pm)
&=
\sqrt{1-\beta^{2}}\,\sin\theta
\left[
e^{2}Q_{\ell}Q_{t}
+
\frac{s\,g_{Z}^{R\ell}}{2}
\frac{g_{Z}^{Lt}+g_{Z}^{Rt}}
{\Delta_Z}
\right] \\
\mathcal{M}_{\rm SM}(-,+;\pm,\pm)
&=
-\sqrt{1-\beta^{2}}\,\sin\theta
\left[
e^{2}Q_{\ell}Q_{t}
+
\frac{s\,g_{Z}^{L\ell}}{2}
\frac{g_{Z}^{Lt}+g_{Z}^{Rt}}
{\Delta_Z}
\right]  \\
\mathcal{M}_{\rm SM}(+,-;\pm,\mp)
&=
(1\pm\cos\theta)
\left[
e^{2}Q_{\ell}Q_{t}
+
\frac{s\,g_{Z}^{R\ell}}{2}
\frac{
g_{Z}^{Lt}(1\mp\beta)
+
g_{Z}^{Rt}(1\pm\beta)
}
{\Delta_Z}
\right]  \\
\mathcal{M}_{\rm SM}(-,+;\pm,\mp)
&=
(1\mp\cos\theta)
\left[
e^{2}Q_{\ell}Q_{t}
+
\frac{s\,g_{Z}^{L\ell}}{2}
\frac{
g_{Z}^{Lt}(1\mp\beta)
+
g_{Z}^{Rt}(1\pm\beta)
}
{\Delta_Z}
\right]
\label{SM-process}
\end{align}
where $Q_\ell=-1$, $Q_t=2/3$, $M_Z=91.2$ GeV, and $\Gamma_Z=2.4955$ GeV \cite{ParticleDataGroup:2024cfk}.  The SM chiral couplings are
\begin{align}
g_Z^{L_{(\ell,t)}}=\frac{e (T_3^{(\ell,t)}-Q_{(\ell, t)} \sin^2\theta_W)}{\cos\theta_W \sin\theta_W},\,\,\,
g_Z^{R_{(\ell,t)}}&=-\frac{e Q_{(\ell,t)} \sin\theta_W}{\cos\theta_W}.
\end{align}
For the numerical analysis we use $\sin^2\theta_W=0.23343$, together with $T_3^\ell=-1/2$ and $T_3^t=1/2$, consistently with Ref.~\cite{Arai:2026jtc}.
%%%%%%%%%%%%%%%%%%%%%%%%%%%%%%%%%%%%%%%%%%%%%%%%%%%%%%%%%%%%%%%%%%%%%%%%%%%%%%%%%%%%%%%%%%%%%%
%\subsection{$Z^\prime$ induced $U(1)_X$ scenario}
%%%%%%%%%%%%%%%%%%%%%%%%%%%%%%%%%%%%%%%%%%%%%%%%%%%%%%%%%%%%%%%%%%%%%%%%%%%%%%%%%%%%%%%%%%%%%%
The $s$-channel $Z^\prime$ helicity amplitudes in the $U(1)_X$ scenario are
\begin{align}
\mathcal{M}_{Z^\prime}(+,-;\pm,\pm)
&=
-\frac{
g_X^2\,m_t\sqrt{s}\,(x_H+1)(5x_H+4)\sin\theta
}{
6\,\Delta_{Z'}
}  \\
%\\[6pt]
\mathcal{M}_{Z^\prime}(+,-;\pm,\mp)
&=
-\frac{
g_X^2\sqrt{s}\,(x_H+1)
\left[
\sqrt{s}(5x_H+4)
\pm 3\sqrt{s-4m_t^2}\,x_H
\right]
(1\pm\cos\theta)
}{
12\,\Delta_{Z'}
}  \\
%\\[8pt]
\mathcal{M}_{Z^\prime}(-,+;\pm,\pm)
&=
\frac{
g_X^2\,m_t\sqrt{s}\,(x_H+2)(5x_H+4)\sin\theta
}{
12\,\Delta_{Z'}
}\\
%\\[6pt]
\mathcal{M}_{Z^\prime}(-,+;\pm,\mp)
&=
-\frac{
g_X^2\sqrt{s}\,(x_H+2)
\left[
\sqrt{s}(5x_H+4)
\pm 3\sqrt{s-4m_t^2}\,x_H
\right]
(1\mp\cos\theta)
}{
24\,\Delta_{Z'}
}
\label{genU1}
\end{align}
where $g_X$ is the general $U(1)_X$ coupling and $\Gamma_{Z^\prime}$ is the total decay width of $Z^\prime$. For $x_H=-2$ and $x_\Phi=1$, the general $U(1)_X$ model reduces to the $U(1)_R$ scenario. The same SM-fermion charge assignment is also obtained from the $U(1)_{xq-\tau_R^3}$ model by choosing $x=0$, while it has no equivalent realization in the $U(1)_{q+\tilde{x}u}$ framework. In this case the left-handed top quark has no direct coupling to the $Z^\prime$ boson. The choice $x_H=-0.5$ in the $U(1)_X$ model, equivalent to $x=1/4$ in $U(1)_{xq-\tau_R^3}$ and $\tilde{x}=0$ in $U(1)_{q+\tilde{x}u}$, gives a complementary chiral pattern in which the right-handed top quark has no direct $Z^\prime$ coupling. The vector-like B$-$L scenario is recovered by taking $x_H=0$ and $x_\Phi=1$ in the $U(1)_X$ model, $x=1/3$ in the $U(1)_{xq-\tau_R^3}$ model, and $\tilde{x}=1$ in the $U(1)_{q+\tilde{x}u}$ model. Finally, the charge assignment $x_H=2$ and $x_\Phi=1$ is equivalent to $x=2/3$ in the $U(1)_{xq-\tau_R^3}$ model and cannot be embedded in the $U(1)_{q+\tilde{x}u}$ framework. These representative charge assignments, summarized in Tab.~\ref{tab1}, are used throughout our analysis to test how the $t\bar t$ spin-density matrix responds to different left- and right-handed $Z^\prime$ couplings.
%%%%%%%%%%%%%%%%%%%%%%%%%%%%%%%%%%%%%%%%%%%%%%%%
\subsection{Limits on model parameters from colliders}
%%%%%%%%%%%%%%%%%%%%%%%%%%%%%%%%%%%%%%%%%%%%%%%%%%
We derive constraints on the $(g_X, M_{Z^\prime})$ plane for different $x_H$ using existing collider data. Limits are obtained from $Z$-pole observables at LEP~\cite{ALEPH:2013dgf}, dilepton and dijet searches at the LHC (CMS~\cite{CMS:2019tbu}, ATLAS~\cite{ATLAS:2019erb}), and contact-interaction bounds from LEP-II~\cite{ALEPH:2013dgf}. The resulting lower limits on $M_{Z^\prime}/g_X$ from LEP-II are shown in Tab.~\ref{tab2}, and the benchmark values used in this analysis are listed in Tab.~\ref{tab3}. For a detailed discussion of the constraint derivation, see \cite{KA:2023dyz,Das:2024gfg}.
%%%%%%%%%%%%%%%%%%%%%%%%%%%%%%%%%%%%%%%%%%%%%%%
\begin{table}[!htbp]
\begin{center}
\begin{tabular}{|c|c|c|c|c|c|}
\hline
\multirow{2}{*}{Machine} & \multirow{2}{*}{$\sqrt{s}$} & \multicolumn{4}{|c|}{95\% CL lower limit on $M_{Z'}/g_X$ (in TeV)} \\ \cline{3-6}
& & $x_H=-2$ &  $x_H=-0.5$ &$x_H=0$ & $x_H=2$ \\ \hline
LEP-II & 209 GeV & 5.0 & 4.4&7.0 &18.0 \\
\hline
\hline
\end{tabular}
\end{center}
\caption{The $95\%$ CL lower limits on $M_{Z^\prime}/g_X$ in the $U(1)_X$ model from $e^+e^-\to f\bar{f}$ processes for different $x_H$.}
\label{tab2}
\end{table}
%%%%%%%%%%%%%%%%%%%%%%%%%%%%%%%%%%%%%%%%%%%%%%%%%%%%%%%%%%%%
\begin{table}[!htbp]
\begin{center}
\begin{tabular}{ |c| c| c|}
\hline
\hline
 $U(1)_X$ charge & $M_{Z^\prime}$& Upper bound on $g_X$ \\
\hline
$x_H=-2$, $x_\Phi=1$& (5 TeV, 7.5 TeV) & (0.16,0.75)\\
\hline
$x_H=-0.5$, $x_\Phi=1$& (5 TeV, 7.5 TeV) & (0.3,0.75)\\
\hline
$x_H=0$, $x_\Phi=1$& (5 TeV, 7.5 TeV) & (0.2,0.75)\\
\hline
$x_H=2$, $x_\Phi=1$& (5 TeV, 7.5 TeV) & (0.06, 0.45)\\
\hline
\hline
\end{tabular}
\end{center}
\caption{Benchmark values of $M_{Z^\prime}$ and strongest upper bounds on $g_X$ for different $x_H$ under the general $U(1)_X$ from LHC and LEP-II experiments. $M_{Z^\prime}=5$ TeV is constrained by LHC dilepton searches; $M_{Z^\prime}=7.5$ TeV is constrained by LEP-II for $x_H=2$ and by LHC otherwise.}
\label{tab3}
\end{table}
%%%%%%%%%%%%%%%%%%%%%%%%%%%%%%%%%%%%%%%%%%%%%%%%%%%%%%%%%%%
%\clearpage
\section{Results and discussion}
\label{RD}
%%%%%%%%%%%%%%%%%%%%%%%%%%%%%%%%%
We now discuss the numerical behavior of the quantum-information observables in the general $U(1)_X$ scenario.  We first consider unpolarized initial beams, corresponding to Eq.~(\ref{spinmat}), and then turn to polarized $e^-e^+$ beams using Eq.~(\ref{spinmat-pol}).  The unpolarized results provide the baseline against which the polarization effects should be interpreted.  Since the observables are constructed from the normalized spin density matrix, they are not simply proxies for the total rate; they probe how the helicity structure of $\gamma/Z/Z^\prime$ exchange reshapes the two-qubit spin state of the final $t\bar t$ pair.

\begin{figure}[!htbp]
\centering
\includegraphics[width=\textwidth]{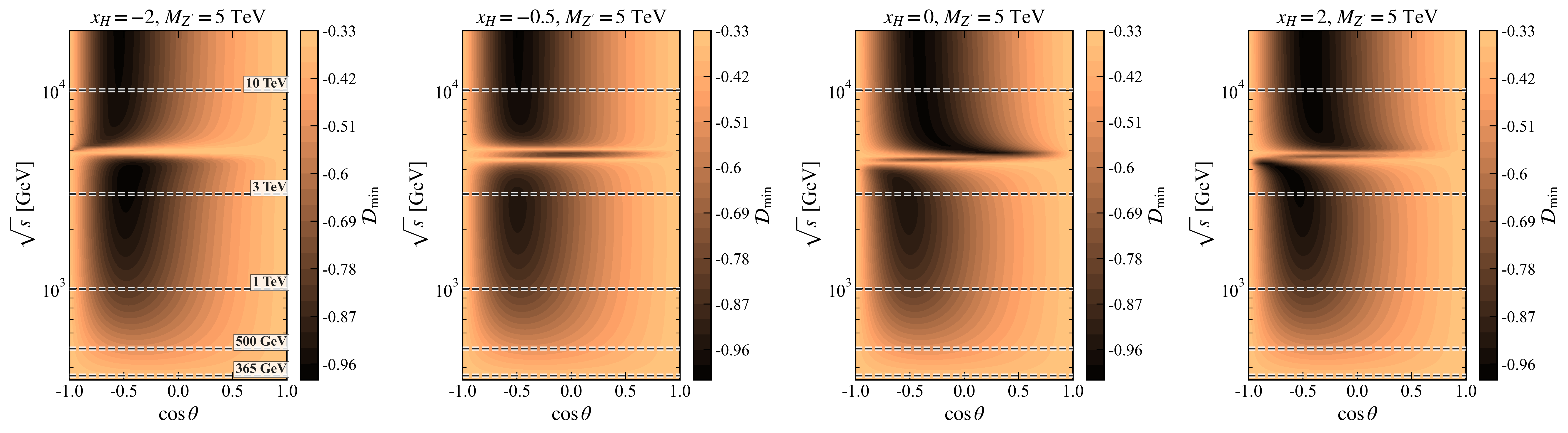}\\[2mm]
\includegraphics[width=\textwidth]{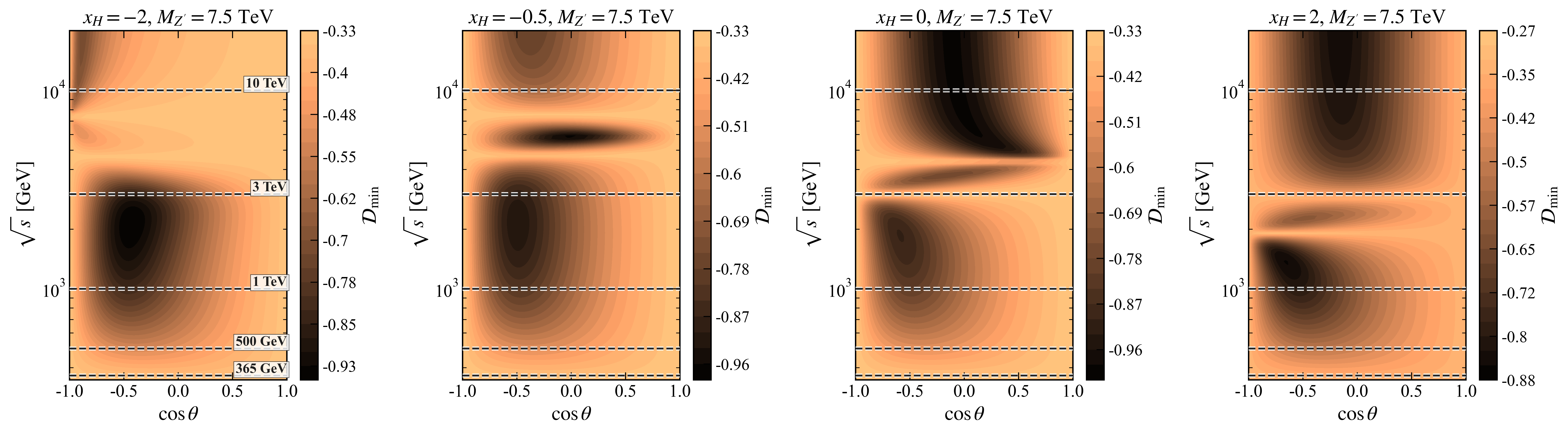}
\caption{Variation of $\mathcal{D}_{\min}$ in the $\sqrt{s}$--$\cos\theta$ plane for the four $x_H$ benchmarks in the $U(1)_X$ scenario.  The upper and lower rows correspond to $M_{Z^\prime}=5$ and $7.5$ TeV, respectively, with the couplings from Tab.~\ref{tab3}.  Black dashed lines with white outlines indicate the representative $e^-e^+$ and $\mu^-\mu^+$ collider energies.}
\label{fig:Dmin}
\end{figure}

Fig.~\ref{fig:Dmin} shows the sufficient entanglement marker $\mathcal{D}_{\min}$ over the $\sqrt{s}$--$\cos\theta$ plane up to $\sqrt{s}=20$ TeV.  The white dashed lines mark representative lepton-collider energies, including the 10 TeV muon-collider benchmark.  Regions with $\mathcal{D}_{\min}<-1/3$ correspond to spin-entangled $t\bar t$ states according to Eq.~(\ref{suff2}).  The pattern depends sensitively on $x_H$ because left- and right-handed SM fermions carry different $U(1)_X$ charges.  The angular dependence of $\mathcal{D}_{\min}$ therefore probes the chiral structure of the $Z^\prime$ interaction in addition to its mass and width.

\begin{figure}[!htbp]
\centering
\includegraphics[width=\textwidth]{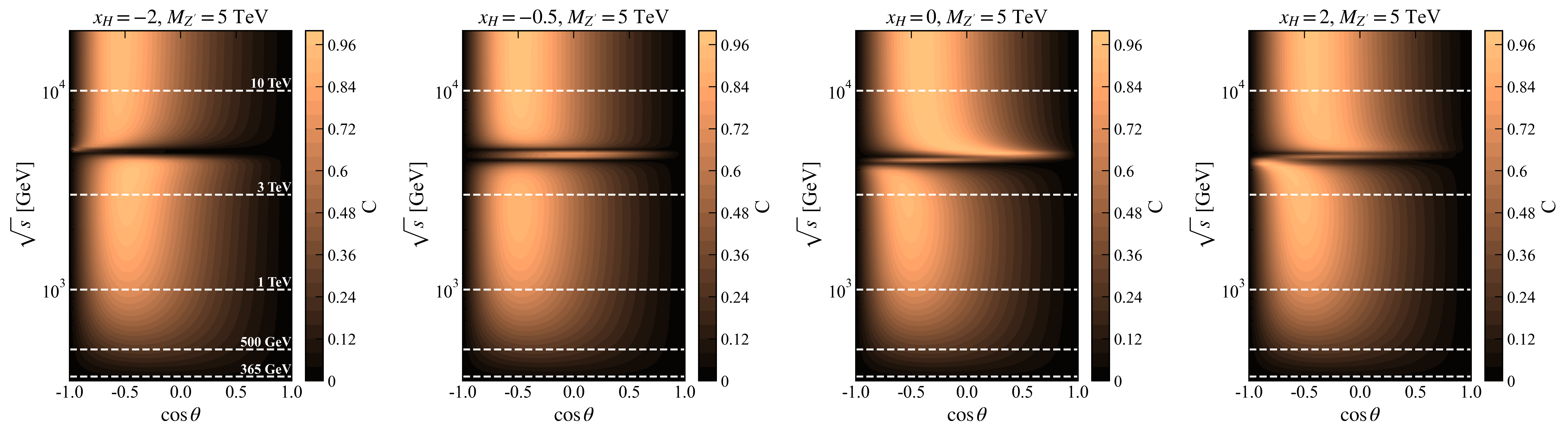}\\[2mm]
\includegraphics[width=\textwidth]{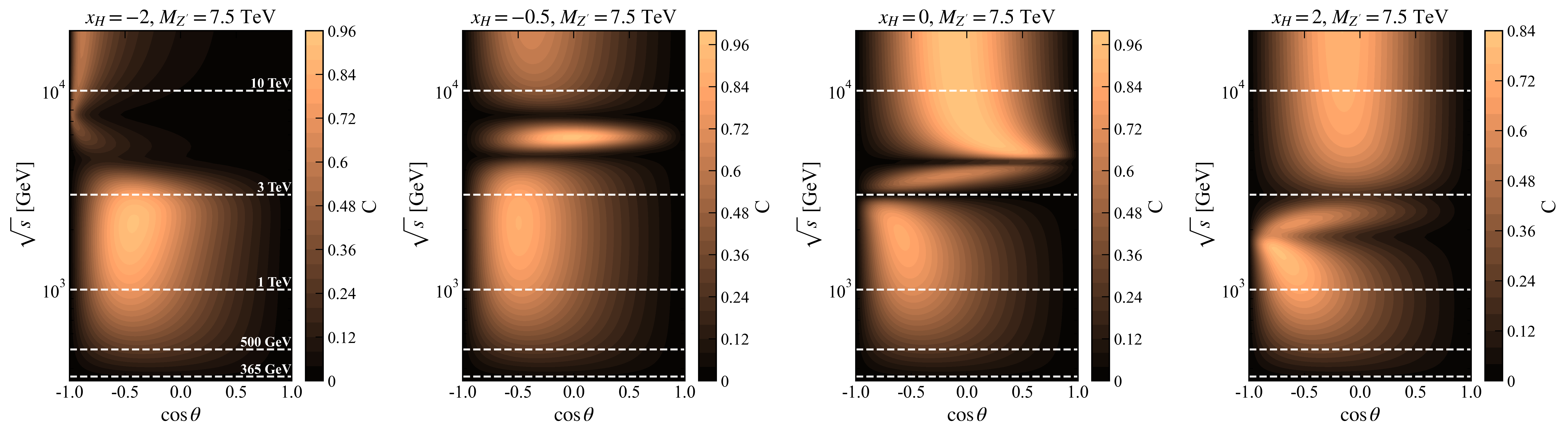}
\caption{Concurrence in the $\sqrt{s}$--$\cos\theta$ plane for the four $x_H$ benchmarks. The upper and lower rows correspond to $M_{Z^\prime}=5$ and $7.5$ TeV, respectively, with the couplings from Tab.~\ref{tab3}. Dashed white lines indicate the representative $e^-e^+$ and $\mu^-\mu^+$ collider energies.}
\label{fig:concurrence}
\end{figure}

The concurrence in Fig.~\ref{fig:concurrence} provides a complementary, basis-independent entanglement measure.  While $\mathcal{D}_{\min}$ is a sufficient marker built from the diagonal spin correlations, concurrence uses the full density matrix.  The agreement of the broad entangled regions in Figs.~\ref{fig:Dmin} and \ref{fig:concurrence} confirms that the diagonal spin correlations capture the dominant entanglement structure in these benchmarks, but the concurrence also reveals where off-diagonal density-matrix elements enhance or reduce the entanglement.

\begin{figure}[!htbp]
\centering
\includegraphics[width=\textwidth]{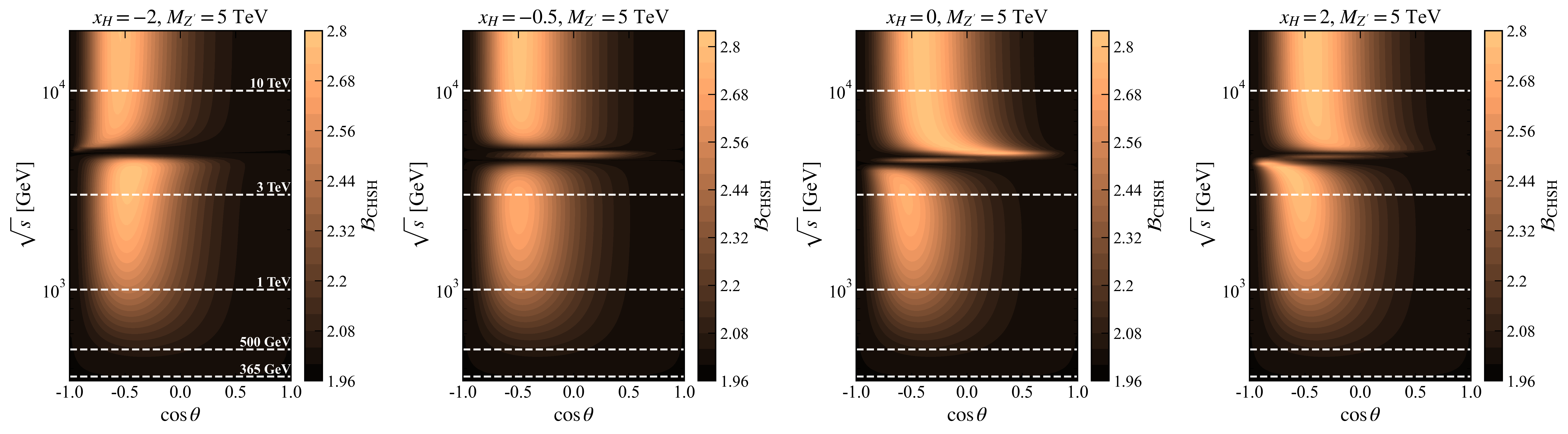}\\[2mm]
\includegraphics[width=\textwidth]{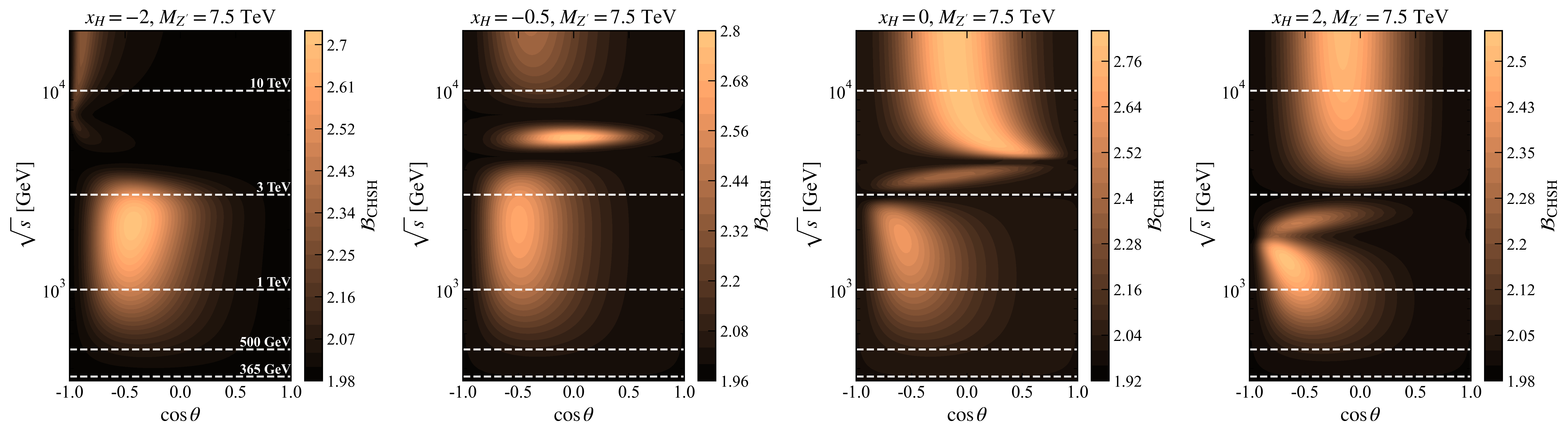}
\caption{Bell-CHSH parameter $\mathcal B_{\rm CHSH}$ in the $\sqrt{s}$--$\cos\theta$ plane for the four $x_H$ benchmarks.  The upper and lower rows correspond to $M_{Z^\prime}=5$ and $7.5$ TeV, respectively, with the couplings from Tab.~\ref{tab3}.  Dashed white lines indicate the representative $e^-e^+$ and $\mu^-\mu^+$ collider energies.}
\label{fig:CHSH}
\end{figure}

Fig.~\ref{fig:CHSH} shows $\mathcal B_{\rm CHSH}$.  The local bound is $\mathcal B_{\rm CHSH}\leq2$, so values above $2$ signal CHSH violation and Bell nonlocality. All states that violate the CHSH inequality are entangled; however, some entangled states satisfy $\mathcal B_{\rm CHSH}\leq2$ and therefore do not exhibit CHSH violation. Consequently, the CHSH plots identify the kinematic windows in which the $U(1)_X$ interaction produces the strongest genuinely nonlocal spin correlations.  In the $x_H=0$ limit our setup reduces to the minimal B$-$L charge assignment, and the SM/B$-$L behavior of the entanglement and CHSH observables agrees with the recent lepton-collider analysis of Ref.~\cite{Arai:2026jtc}.  This provides a useful cross-check before moving to the more chiral $x_H\neq 0$ benchmarks.

\begin{figure}[!htbp]
\centering
\includegraphics[width=0.83\textwidth]{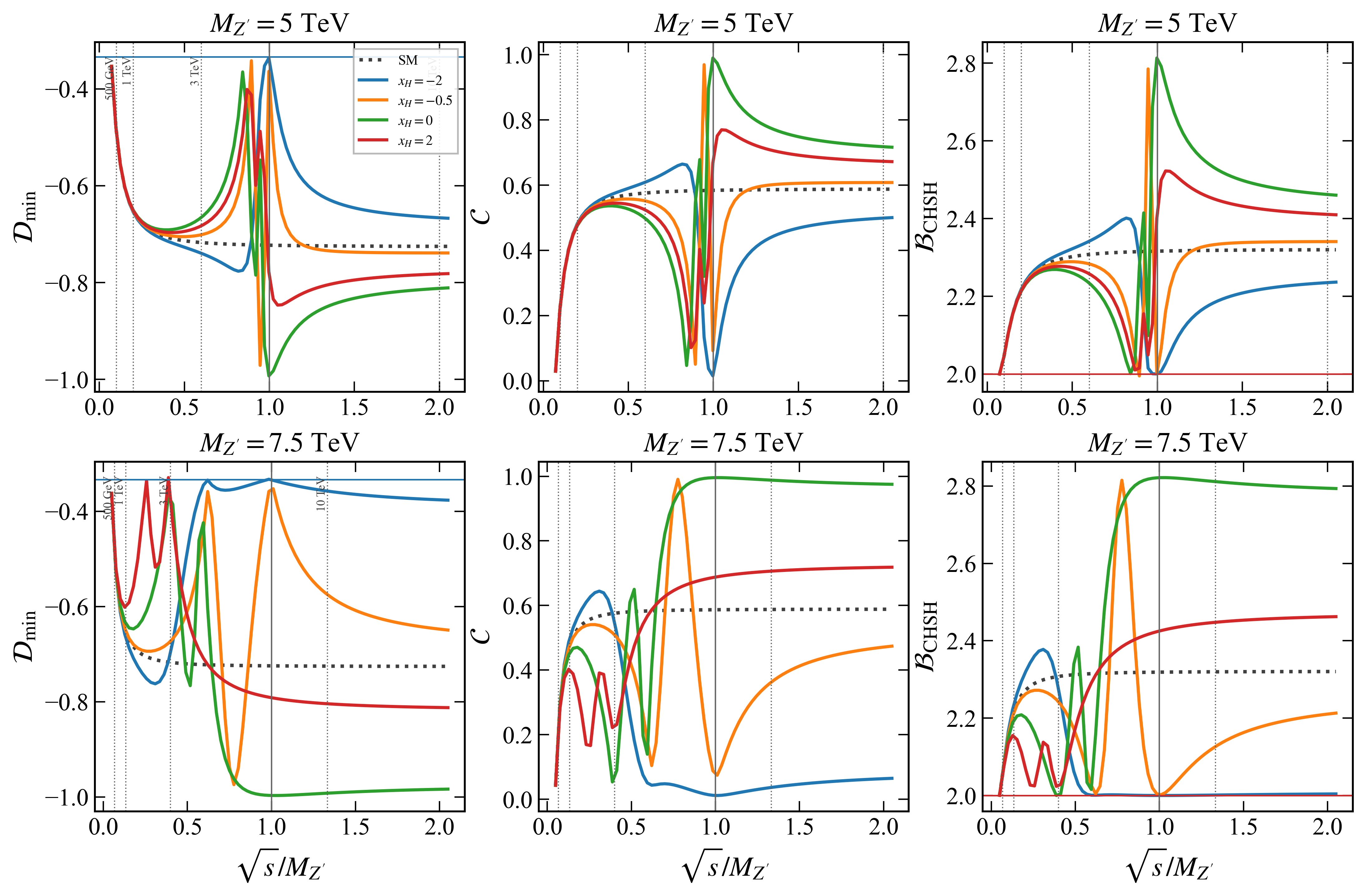}
\caption{Dependence of $\mathcal D_{\min}$ (left), concurrence (middle), and the maximal CHSH parameter (right) on $\sqrt{s}/M_{Z^\prime}$ at $\cos\theta=0$.  The upper and lower rows correspond to $M_{Z^\prime}=5$ and $7.5$ TeV, respectively, with the couplings from Tab.~\ref{tab3}.  Vertical dotted lines indicate representative lepton-collider energies.}
\label{fig:qi-smzp}
\end{figure}

The fixed-angle scan in Fig.~\ref{fig:qi-smzp} makes the energy dependence relative to the $Z^\prime$ pole more transparent.  At $\theta=\pi/2$, the observables vary sharply as $\sqrt{s}$ approaches the $Z^\prime$ mass, but the size and sign of the deformation depend on $x_H$.  This behavior motivates the polarized $e^-e^+$ study below: electron and positron beam polarization can select the $e^-_R e^+_L$ or $e^-_L e^+_R$ helicity channel, giving direct access to the charge combinations that are averaged together in the unpolarized plots.

\subsection{Polarized $e^-e^+$ beam effects}

\begin{figure}[!htbp]
\centering
\includegraphics[width=0.98\textwidth]{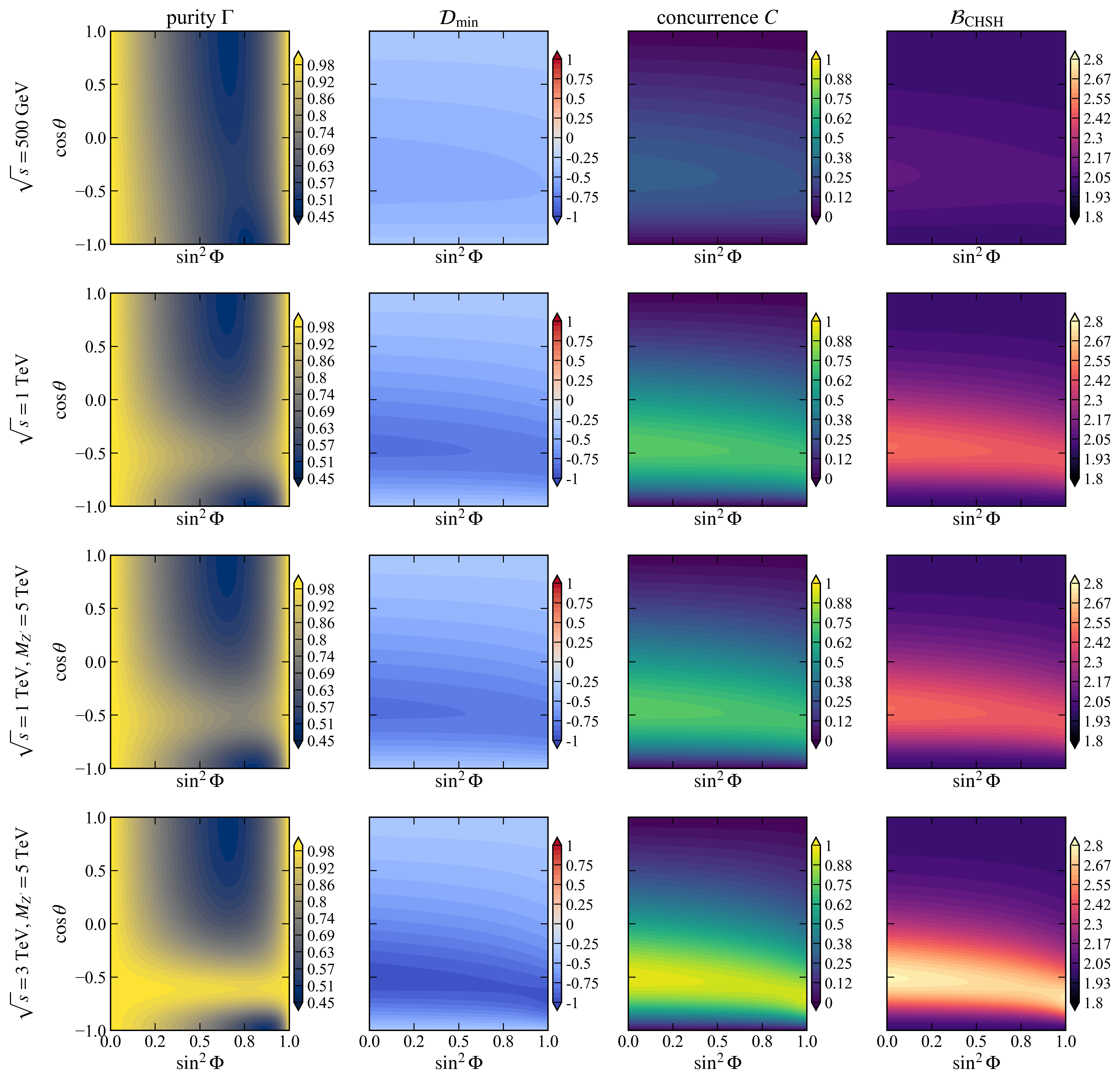}
\caption{Polarized-beam spin-state landscapes for $e^-e^+\to t\bar t$.  The columns show the purity $\Gamma$, $\mathcal{D}_{\min}$, concurrence, and $\mathcal B_{\rm CHSH}$ in the $\sin^2\Phi$--$\cos\theta$ plane.  The first two rows show the SM at $500$ GeV and $1$ TeV.  The last two rows show the $U(1)_X$ benchmark with $x_H=0$ and $M_{Z^\prime}=5$ TeV at $\sqrt{s}=1$ and $3$ TeV.}
\label{fig:pol-landscape}
\end{figure}

The preceding figures average the two allowed initial helicity channels with equal weights. Future linear $e^-e^+$ colliders such as ILC and CLIC can instead prepare polarized beams, making the mixture in Eq.~(\ref{sinphi}) experimentally tunable. We do not assume an analogous polarization program for future muon colliders. This opens up a new way to use quantum observables in the electron-collider case: one may compare the spin density matrices obtained from different helicity mixtures.  As a validation of the implementation, we reproduced the known polarized-beam SM behavior of $e^-e^+\to t\bar t$ \cite{Altakach:2026fpl}; the corresponding SM panels are included in the first two rows of Fig.~\ref{fig:pol-landscape}. For the SM validation, we find $\max C=0.297$ and $\max \mathcal B_{\rm CHSH}=2.086$ at $\sqrt{s}=500$ GeV, and $\max C=0.742$ and $\max \mathcal B_{\rm CHSH}=2.491$ at $\sqrt{s}=1$ TeV. These values reproduce the corresponding published SM polarized-beam results within the expected difference from input constants and grid resolution. The important physics lesson is that beam polarization need not induce a large change in normalized spin observables if the two helicity channels lead to similar final-state density matrices. The rate can change substantially while concurrence and CHSH remain relatively stable.

Fig.~\ref{fig:pol-landscape} compares the SM and $U(1)_X$ landscapes. At $500$ GeV, the SM concurrence and $\mathcal B_{\rm CHSH}$ are nearly independent of both $\cos \theta$ and the effective polarization angle, whereas at $1$ TeV the top quarks are more relativistic and the polarization dependence becomes more visible. The $U(1)_X$ benchmark shows a stronger response at the multi-TeV electron-collider benchmark $\sqrt{s}=3$ TeV. In this region the two initial helicity channels probe different combinations of $q_{\ell_L}$ and $q_{\ell_R}$, producing distinct patterns in $\Gamma$, $\mathcal{D}_{\min}$, concurrence, and $\mathcal B_{\rm CHSH}$.

\begin{figure}[!htbp]
\centering
\includegraphics[width=0.9\textwidth]{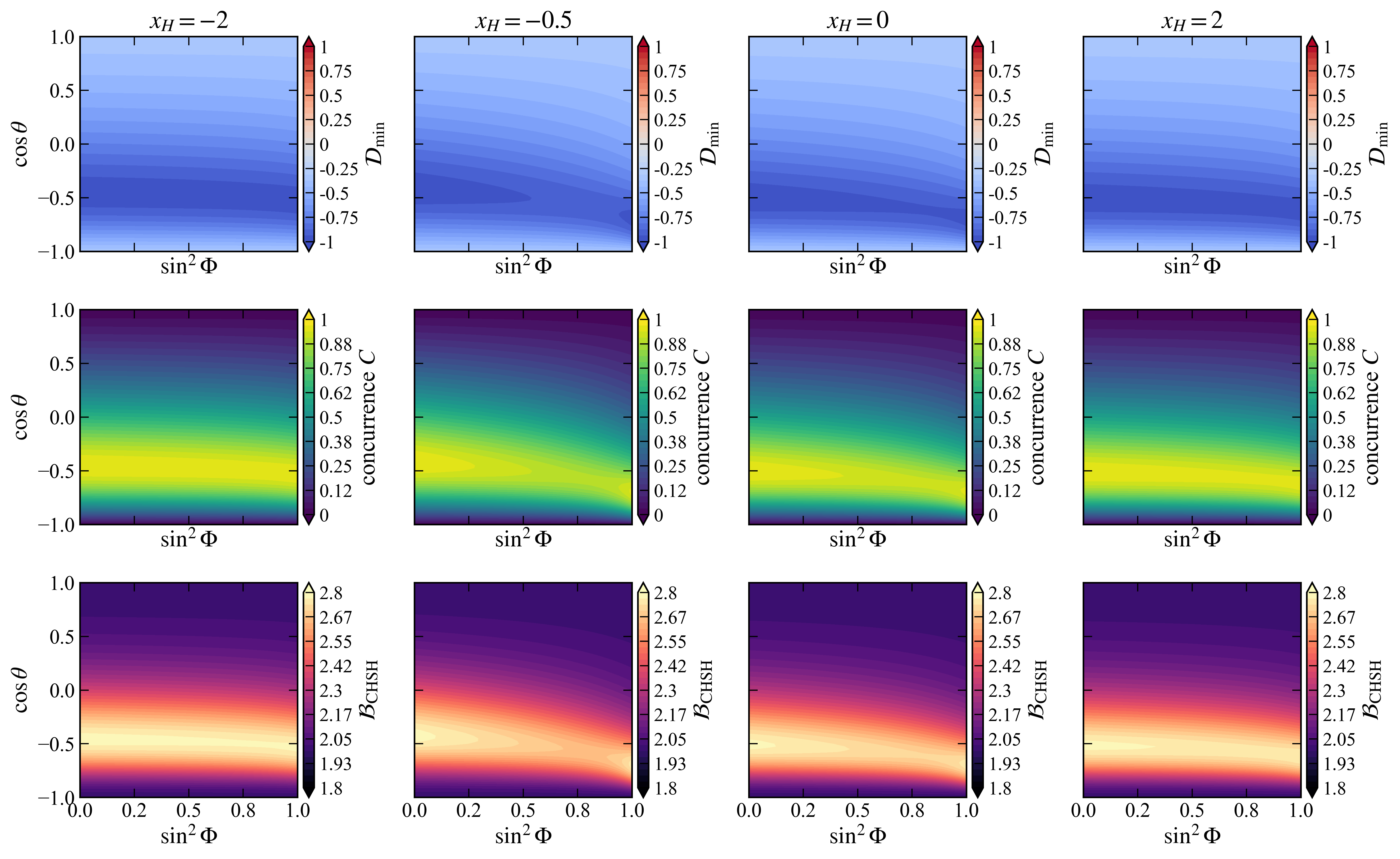}
\caption{$U(1)_X$ polarization response at $\sqrt{s}=3$ TeV for $M_{Z^\prime}=5$ TeV.  The columns correspond to representative $x_H$ values, and the rows show $\mathcal D_{\min}$, concurrence, and $\mathcal B_{\rm CHSH}$.  Beam polarization separates the $q_{\ell_R}$- and $q_{\ell_L}$-weighted amplitudes, making the $x_H$ dependence more visible than in the unpolarized average.}
\label{fig:pol-xh-scan}
\end{figure}

The $x_H$ scan in Fig.~\ref{fig:pol-xh-scan} illustrates how polarization turns the charge assignment into a directly observable spin-pattern effect.  We keep $M_{Z^\prime}=5$ TeV and $\sqrt{s}=3$ TeV fixed, using the benchmark couplings in Tab.~\ref{tab3}.  The $U(1)_X$ charges $q_{\ell_R}=-x_H-1$ and $q_{\ell_L}=-x_H/2-1$ are affected differently by varying $x_H$, so the right- and left-handed initial channels need not generate the same normalized spin state.  Showing $\mathcal D_{\min}$ together with concurrence tests whether the diagonal-correlation criterion tracks the entanglement obtained from the full density matrix as the helicity mixture is varied.

\begin{figure}[!htbp]
\centering
\includegraphics[width=0.9\textwidth]{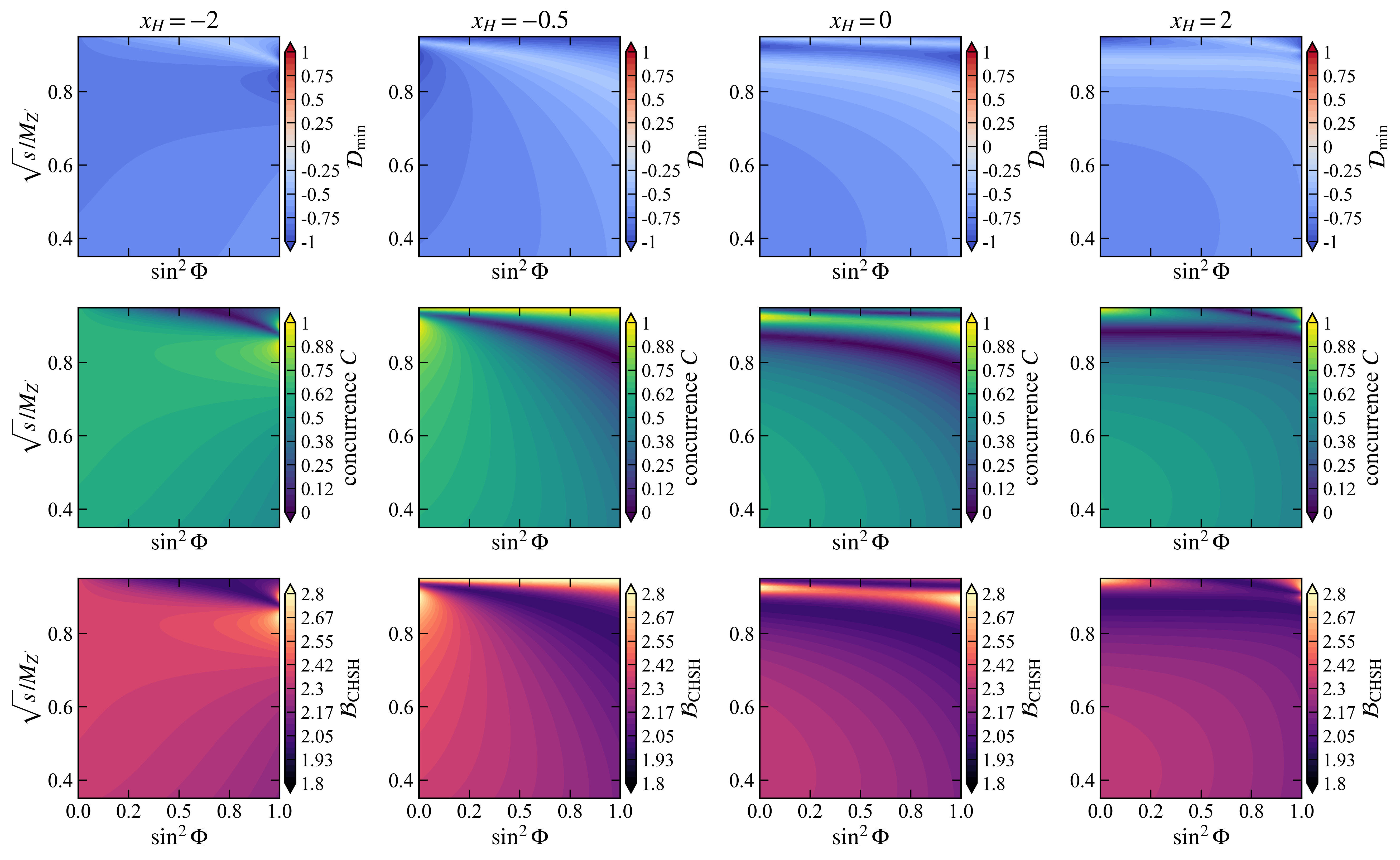}
\caption{Polarized \(U(1)_X\) response as a function of \(\sqrt{s}/M_{Z^\prime}\) and \(\sin^2\Phi\) at fixed \(M_{Z^\prime}=5\) TeV and \(\cos\theta=0\).  The columns show representative \(x_H\) choices and the rows show \(\mathcal D_{\min}\), concurrence, and \(\mathcal B_{\rm CHSH}\).  This plot isolates the resonance-proximity effect that controls how strongly the polarized observables separate different chiral charge assignments.}
\label{fig:pol-ratio-scan}
\end{figure}

Although Fig.~\ref{fig:pol-xh-scan} illustrates the dependence on \(x_H\), the differences among the charge assignments are less pronounced away from the \(Z^\prime\) resonance. To make the role of resonance proximity explicit, Fig.~\ref{fig:pol-ratio-scan} shows the polarized observables as functions of \(\sqrt{s}/M_{Z^\prime}\), with \(M_{Z^\prime}=5\) TeV and \(\cos\theta=0\). As \(\sqrt{s}\) approaches \(M_{Z^\prime}\), the enhanced \(Z^\prime\) contribution and its interference with the SM amplitudes increase the separation among the different chiral charge assignments. Far below the resonance, the \(Z^\prime\) exchange is well approximated by a contact interaction and remains subdominant, reducing the sensitivity of the normalized spin observables to \(x_H\).

\begin{figure}[!htbp]
\centering
\includegraphics[width=0.98\textwidth]{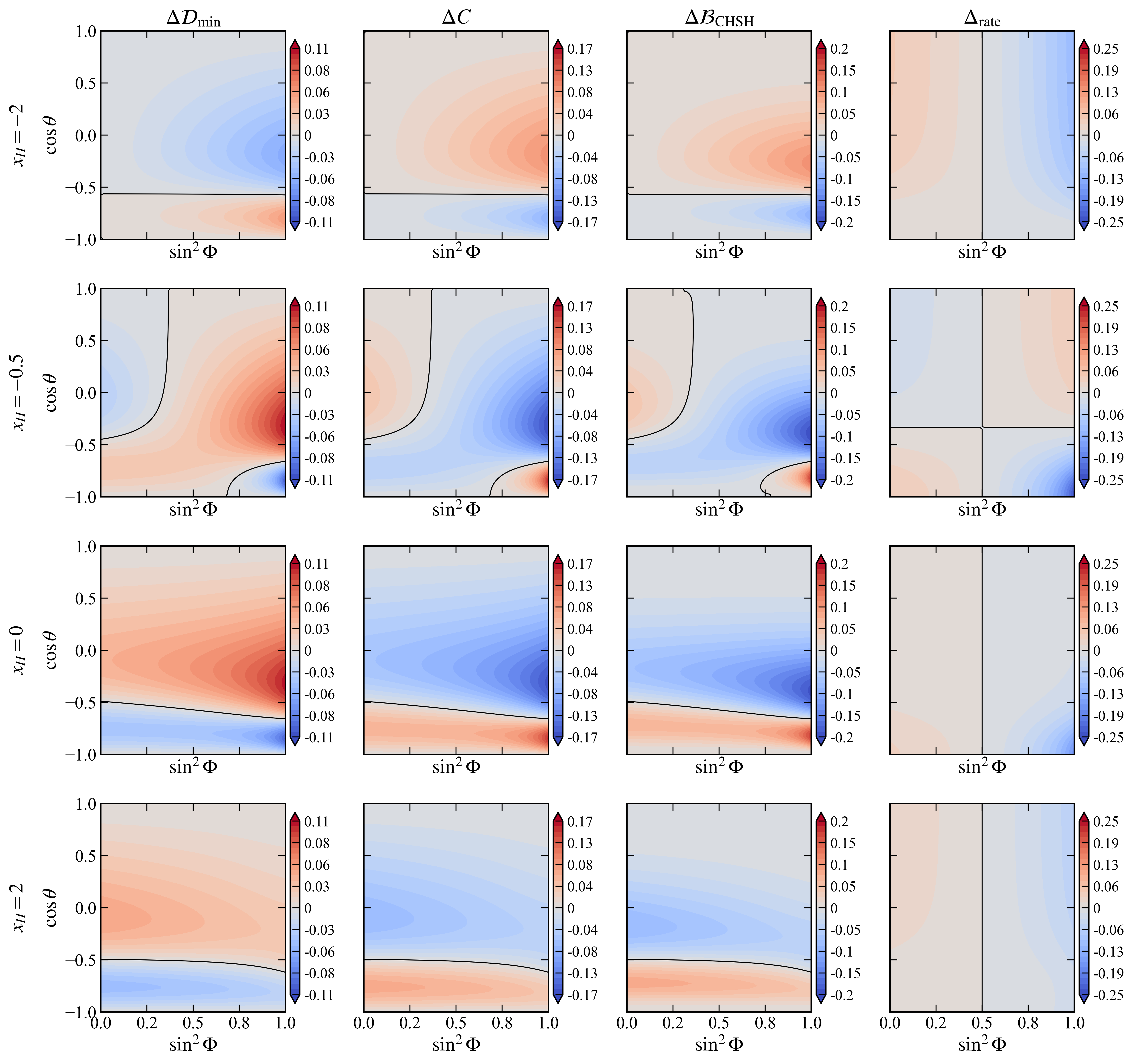}
\caption{Polarized deviations at $\sqrt{s}=3$ TeV and $M_{Z^\prime}=5$ TeV.  Rows correspond to $x_H=-2,-0.5,0,2$, and columns show $\Delta\mathcal{D}_{\min}$, $\Delta C$, $\Delta\mathcal B_{\rm CHSH}$ and $\Delta_{\rm rate}$, following Eq.~(\ref{delta-defs}).  Each column uses a common color range across the four charge assignments.  Black zero contours separate enhancements from suppressions relative to the SM.}
\label{fig:pol-u1x-minus-sm}
\end{figure}

Finally, Fig.~\ref{fig:pol-u1x-minus-sm} compares the polarized deviations for all four charge benchmarks using the definitions in Eq.~(\ref{delta-defs}).  A common color range is used down each observable column, allowing the magnitude and sign of the shifts to be compared directly among the charge assignments.  Unlike an individual polarized cross section, the double ratio $\Delta_{\rm rate}$ is fixed by $\sin^2\Phi$ because the common factor $W_{RL}+W_{LR}$ cancels between the $U(1)_X$ and SM polarization ratios.  Its largest shift does not always coincide with the largest shift in $\mathcal{D}_{\min}$, concurrence, or $\mathcal B_{\rm CHSH}$.  This separation is useful phenomenologically: spin-density-matrix observables provide information complementary to cross sections and can help distinguish chiral $Z^\prime$ interactions that would otherwise give similar unpolarized rates.

%%%%%%%%%%%%%%%%%%%%%%%%%%%%%%%%%%%%%%%%%%%%%%%%%%%%%%%%%%%%

%%%%%%%%%%%%%%%%%%%%%%%%%%
\clearpage
\section{Conclusions}
\label{sec:conc}
%%%%%%%%%%%%%%%%%%%%%%%%%%%%%%

We have studied the spin structure of $\ell^-\ell^+\to t\bar t$ in anomaly-free general $U(1)$ extensions of the SM using the full production density matrix, including the interference of photon, $Z$, and $Z^\prime$ exchange. By varying $x_H$ over representative $U(1)_X$ benchmarks, we find that the entanglement marker $\mathcal D_{\min}$, concurrence, purity, and maximal CHSH parameter respond differently to the resulting chiral charge assignments and therefore provide information complementary to differential rates, with the largest distinctions appearing as $\sqrt{s}$ approaches the $Z^\prime$ pole.  For polarized $e^-e^+$ beams, the effective polarization angle $\Phi$ continuously selects the two opposite-helicity initial channels and exposes their different left- and right-handed $Z^\prime$ couplings; at the representative $3$ TeV electron-collider benchmark this produces charge-dependent spin patterns that can remain visible even when rate modifications alone are less diagnostic.  The $x_H=0$ benchmark reproduces the vector-like B$-$L limit and agrees with independent SM and B$-$L studies, while the polarized SM calculation provides a further cross-check of the implementation.  A realistic assessment of this program will require top-decay spin analyzers, detector effects, event selection, and statistical uncertainties, but the present results establish quantum spin observables as a useful additional probe of chiral neutral currents at future lepton colliders.

%%%%%%%%%%%%%%%%%%%%%%%%%%%%%%%%%%%%%%%%%%%%%%%%%%%%%%%%%%%%%%%%%%%%%%%
\black
%%%%%%%%%%%%%%%%%%%%%%%%%%%%%%%%%%%%%%%%%%%%%%%%%%%%%%%%%%%%%%%%%%%%
\begin{acknowledgments}
S.K.A. is supported by JST SPRING, Grant Number JPMJSP2119.  The work of S.M. is supported by a KIAS Individual Grant (PG086002) at the Korea Institute for Advanced Study.  A.D. thanks Osaka University for its hospitality during HPNP2025. A. D. also thanks Dibyashree Sengupta and Amit Kumar Pal for useful discussions. S.K.A. thanks Valentin Durupt for providing the MadGraph density-matrix implementation.
%%%%%%%%%%%%%%%%%%%%%%%%%%%%%%%%%%%
\end{acknowledgments}
%%%%%%%%%%%%%%%%%%%
\bibliographystyle{utphys}
\bibliography{bibliography}
\end{document}